\documentclass[review]{elsarticle}
\usepackage{amsmath}
\usepackage{lineno}
\usepackage[backref=page]{hyperref}
\usepackage{caption,setspace}
\usepackage{subcaption}
\usepackage{algpseudocode}
\usepackage{float}
\usepackage{url}
\usepackage[utf8]{inputenc}
\usepackage{multirow}
\usepackage{comment}
\modulolinenumbers[5]
\usepackage{soul,color}
\usepackage{amsfonts}
\usepackage{graphicx}
\journal{ISPRS J. Photogramm}

\begin{document}

\begin{frontmatter}

\title{Inferring 3D change detection \protect \\from bitemporal optical images}

\author[diagaddress]{Valerio Marsocci\corref{mycorrespondingauthor}}
\cortext[mycorrespondingauthor]{Corresponding author: valerio.marsocci@uniroma1.it}

\author[diceaaddress]{Virginia Coletta}
\ead{virginia.coletta@uniroma1.it}

\author[diceaaddress]{Roberta Ravanelli}
\ead{roberta.ravanelli@uniroma1.it}

\author[dietaddress]{\protect \\Simone Scardapane}
\ead{simone.scardapane@uniroma1.it}

\author[diceaaddress,ssas]{Mattia Crespi}
\ead{mattia.crespi@uniroma1.it}

\address[diagaddress]{Department of Computer, Control and Management Engineering Antonio Ruberti (DIAG), Sapienza University of Rome, 00185, Rome, Italy}
\address[diceaaddress]{Department of Civil, Constructional and Environmental Engineering (DICEA), Sapienza University of Rome}
\address[dietaddress]{Department of Information Engineering, Electronics and Telecommunication (DIET), Sapienza University of Rome, 00184, Rome, Italy}
\address[ssas]{Sapienza School for Advanced Studies (SSAS), Sapienza University of Rome, Rome, Italy}

\begin{abstract}
{In recent years, change detection (CD) using deep learning (DL) algorithms has been a very active research topic in the field of remote sensing (RS). Nevertheless, the CD algorithms developed so far are mainly focused on generating two-dimensional (2D) change maps where the planimetric extent of the areas affected by changes is identified without providing any information on the corresponding elevation variations.
The aim of this work is, hence, to establish the basis for the development of DL algorithms able to automatically generate an elevation (3D) CD map along with a standard 2D CD map, using only bitemporal optical images as input, and thus without the need to rely directly on elevation data during the inference phase. 
Specifically, our work proposes a novel network, capable of solving the 2D and 3D CD tasks simultaneously, and a {modified} version of the 3DCD dataset, a freely available dataset designed precisely for this {twofold} task. 
The proposed architecture consists of a {Transformer} network based on a semantic tokenizer: the MultiTask Bitemporal Images {Transformer} (MTBIT).
Encouraging results, obtained on the {modified} version of the 3DCD dataset by comparing the proposed architecture with other networks specifically designed to solve the 2D CD task, are shown. In particular, MTBIT {achieves a metric accuracy (represented by the} changed root mean squared error) equal to 6.46 m -- the best performance among the compared architectures -- with a limited number of parameters (13,1 M).
The code and the 3DCD dataset are available at \url{https://sites.google.com/uniroma1.it/3dchangedetection/home-page}.}
\end{abstract}


\end{frontmatter}


\section{Introduction}
\label{sec:intro}
Change detection (CD) is one of the most active research areas in Remote Sensing (RS) \cite{si2020remote,nascetti_SAR_optical_2D}. 
Specifically, the term CD refers to the process of identifying areas of the Earth’s surface that have experienced changes through the joint analysis of two or more co-registered images captured at different epochs \cite{bruzzone2012novel, daudt2019multitask, si2020remote, 3DCD_nostro}.

CD algorithms, hence, estimate the changes that occurred at ground level and, for this reason, are applied to a wide range of applications, such as natural disaster management \cite{chen2018cities, ZHENG2021112636}, identification of urban changes \cite{lyu2018city,huang2013building, WANG2022112993, HUANG2020111802}, and crops and forests management \cite{khan2017forest}. 
Nowadays, several algorithms can meet the increasingly stringent requirements of CD applications, driven by the unprecedented technological development of Earth Observation (EO) sensors. 
Despite these promising results, CD algorithms are generally focused exclusively on the generation of 2D change maps and thus they are only capable of identifying the areas affected by temporal variations of land use and land cover (LULC) \cite{HERMOSILLA2022112780, goldblatt2018using}, estimating their planimetric (2D) extent. 
In this scenario, deep learning (DL) has overtaken traditional methods like, e.g., change vector analysis (CVA) \cite{cva} and image differencing, proving to be an effective solution to tasks such as semantic segmentation, and object and change detection \cite{deep_learning_RS_review,deep_learning_RS_review2}. 
In particular, several DL architectures are currently employed for solving CD tasks, such as CDNet \cite{alcantarilla2018street}, Siamese models \cite{daudt2018fully}, or ChangeFormer \cite{changeformer} networks.
Nevertheless, the number of CD labelled RS datasets openly available is still limited, since the production of hundreds of annotated images requires a lot of time and specialised knowledge \cite{marsocci2021mare, GUO2021112589}. The scarce availability of RS datasets represents thus an obstacle to the implementation of new methods and limits the development of viable solutions to the aforementioned CD tasks \cite{daudt2019multitask, 3DCD_nostro}.
Moreover, to the best of our knowledge \cite{3D_CD_review,deep_learning_RS_review,deep_learning_RS_review2}, there is no freely available dataset providing pairs of bitemporal optical images and pixel-wise labels identifying both the 2D changes and the corresponding elevation (3D) variations.
For this reason, we specifically designed and produced the 3DCD dataset\footnote{A preliminary version of the dataset was presented in \cite{3DCD_nostro}.} (Section \ref{sec:dataset}), a freely available dataset \cite{website_nostro} containing all the necessary information for solving both the 2D and the 3D CD tasks, starting from optical images captured at different epochs.

One interesting line of research of DL applied to RS is moving precisely towards this direction: to retrieve 3D information from the smallest possible amount of 2D information, usually extracted from optical images \cite{mou2018im2height}. 
Indeed, 2D CD algorithms can only detect planimetric changes such as appearing/disappearing buildings/trees, shrinking/expanding structures; these results, though, are not sufficient to fulfil the requirements of applications needing also the volumetric/vertical information, such as quantitative estimation of {changes in urban areas, forest biomasses and land morphology}  \cite{3D_CD_review,3DCD_nostro}.
The development of suitable 3D CD algorithms can hence offer numerous advantages. We provide a complete overview of the literature on 3D CD and possible applications later on in Section \ref{sec:rel_works}.

\subsection{Contribution of the paper}
The existing state-of-the-art DL CD methods rely mainly on deep CNNs thanks to their ability to extract powerful discriminative features \cite{he2016deep}.
Lately, though, the astounding results of {Transformer}s in Natural Language Processing (NLP) has led researchers in applying {Transformer}s also to computer vision tasks such as image classification and image segmentation, with {architectures like} Vision {Transformer} (ViT) \cite{vit} and SEgmentation {Transformer} (SETR) \cite{setr}, and more recently, to 2D CD with {the} ChangeFormer \cite{changeformer} {model}.
Our work goes one step further by proposing a novel Transformer-based network, trained on the modified version of the recently proposed 3DCD dataset, able to solve simultaneously the 2D and 3D CD tasks, as described in Section \ref{sec:meth}.

Indeed, the main contribution behind our work is to lay the foundations for the development of DL algorithms able to automatically produce 3D CD maps, i.e. maps containing the quantitative changes in elevation – together with standard 2D CD maps, i.e. maps identifying the planimetric extent of the areas affected by elevation changes –, using two optical images acquired in different epochs as
input, without the need of 3D information during the inference step.
In this contribution, we {show} how {the proposed {Transformer}-based architecture} performe{s} very well {when compared with the most recent networks currently available in the literature}, as highlighted in {Section \ref{sec:exp_res}}.
\subsection{Organization of the paper}
The rest of this paper is organized as follows. Section \ref{sec:rel_works} introduces the datasets and the methodologies currently available for the 2D and the 3D CD tasks.
Sections \ref{sec:meth} and \ref{sec:dataset} describe the proposed network and {the modified version of the 3DCD dataset}, respectively.
Section \ref{sec:setup} illustrates the experimental setup used to implement and train the proposed network. 
In Section \ref{sec:exp_res}, we discuss the results, comparing quantitatively and qualitatively the proposed network with {different methods available in literature, designed specifically for 2D CD and adapted to solve also the 3D CD task, and with a baseline, that makes use of the IM2HEIGHT \cite{mou2018im2height} {architecture}. In addition, Section \ref{sec:exp_res} describes also the ablation studies performed to gain a better understanding of the proposed network}.
Finally, we draw conclusions and outlines for future {developments} (Section \ref{sec:conclusions}).

\section{Related works}
\label{sec:rel_works}

In this section we review both the open{ly available} RS CD datasets (Section \ref{sec:rel_datasets}) and the different approaches developed so far to solve the 2D and 3D CD tasks using DL algorithms (Section \ref{sec:rel_meth}). 

\subsection{Related datasets}
\label{sec:rel_datasets}
\sloppy
In general, many datasets that contain {bitemporal} optical images and {pixel-wise 2D CD maps} are currently openly available, while only a small minority include 3D data, namely LiDAR/photogrammetric point cloud and/or Digital Surface Models (DSMs) \cite{3D_CD_review, 3DCD_nostro, lidar_CD, RemoteSensing_3d_CD}.
However, to our knowledge, there {is} no dataset including both {bitemporal} optical images, 2D CD maps and 3D CD maps (Table \ref{tab:data}). 

For this reason, we specifically designed and produced the 3DCD dataset {\cite{3DCD_nostro}}, a freely available dataset providing all the necessary information for solving both the 2D and the 3D CD tasks, starting from bitemporal optical images.
Some details about the production {and the modification} of the proposed dataset are given in Section \ref{sec:dataset}. 

In Table \ref{tab:data}, the main features of the most used CD datasets freely available in literature are recalled.

\begin{table}[h!]
\centering
\caption{Main features of the most used CD datasets freely available in literature.}
\resizebox*{1\textwidth}{!}{
\begin{tabular}{|c|c|c|c|c|}
\hline
Name & N. images & Tile size & CD map & Notes \\ \hline
SZTAKI Air change \cite{benedek2008mixed, benedek2009change} & 13 & 952x640  & 2D &  \\ \hline
DSIFN-CD \cite{zhang2020deeply} & 3600 & 512x512 &  2D &\\ \hline
SECOND \cite{yang2021asymmetric} & 4662 & 512x512 & 2D & 6 classes \\ \hline
OSCD \cite{daudt2018urban} & 24 & 600x600 & 2D & multispectral \\ \hline
S2Looking \cite{shen2021s2looking} & 5000 & 1024x1024 &2D &  \\ \hline
S2MTCP \cite{leenstra2021self} & 1520 & 600x600 &2D &  \\ \hline
SYSU-CD \cite{shi2021deeply} & 20000 & 256x256 &  2D & 6 classes \\ \hline
DynamicEarthNet \cite{dynamicearthnet} & 54750 & 1024x1024 & 2D & dense annotations \\ \hline
LEVIR-CD \cite{chen2020spatial} & 637 & 1024x1024  &  2D &\\ \hline
\textbf{3DCD} \cite{3DCD_nostro} \textbf{(ours)} & 472 & 400x400  &  2D - 3D &\\ \hline
\end{tabular}
}
\label{tab:data}
\end{table}

\subsection{Related methodologies}
\label{sec:rel_meth}
CD is a challenging area of research in RS \cite{pan2022self}. In this section we provide an overview of the available literature on 2D and 3D CD.

\subsubsection{2D Change Detection}

In recent years, several approaches were developed to solve the 2D CD task using deep learning algorithms based on supervised learning, semi-supervised learning, weakly supervised learning, and unsupervised
learning{, including, among the others, CNNs, recurrent neural networks (RNNs), recurrent Boltzmann machines (RBM), Auto-Encoders (AEs), generative adversarial networks (GANs) and {Transformer}-based networks} \cite{pan2022self}.
Moreover, different kinds of imagery (optical, synthetic-aperture-radar (SAR), multispectral{, ...}) have been used to detect 2D changes.
Nevertheless, the majority of urban CD deep learning approaches currently available exclusively employ optical imagery \cite{nascetti_SAR_optical_2D}. Among these,
the three models proposed in \cite{daudt2018fully}, i.e. Fully Convolutional Early Fusion (FC-EF), Fully Convolutional Siamese - Concatenation (FC-Siam-conc), Fully Convolutional Siamese - Difference (FC-Siam-diff),  are among the first forerunners. 
Particuarly, FC-EF processes the two bitemporal images together, concatenating them along their bands, with a U-Net \cite{unet} architecture. Fully Convolutional Siamese (FC-Siam) models, on the other hand, process the input images separately through the encoder. Then the features are concatenated and processed in an unique decoder. The difference between the two aforementioned models relies on the design of the skip connection layer: FC-siam-conc concatenates the two skip connections during the decoding steps, each one coming from one encoding stream, whereas FC-siam-diff concatenates the absolute value of their difference. 

More recent algorithms propose novel approaches. For example, SNUNet-CD (a combination of Siamese network and NestedUNet) \cite{snunet} optimizes the information transmission between encoder and decoder, and between decoder and decoder, through a compact densely connected siamese network and the Ensemble Channel Attention Module (ECAM).

An end-to-end superpixel-enhanced CD network (ESCNet) is proposed in \cite{zhang2021escnet}. Specifically, two weight-sharing superpixel sampling networks (SSNs) encode the features of the image pairs. Then, an UNet-based Siamese network is set as decoder. To enhance the performance, the architecture is integrated with an adaptive superpixel merging (ASM) module and a pixel-level refinement module.

Based on generative super resolution, \cite{liu2021super} proposes a super-resolution-based CD network (SRCDNet), in order to overcome the issue of different resolutions between bitemporal image pairs. Moreover, a stacked attention module (SAM) is presented to extract finer features.

Following a non-supervised approach, \cite{wu2021unsupervised} proposes an unsupervised deep siamese kernel PCA convolutional mapping network (KPCA-MNet). In the KPCA-MNet, the feature encoder is a siamese network made of weight-shared KPCA convolutional layers. Then, the extracted features are mapped into a 2D polar domain. Finally, the CD results are generated by threshold segmentation and clustering, not requiring any labels.

Recently, the ChangeFormer \cite{changeformer} architecture was proposed. Based on a bitemporal image {Transformer} (BIT) \cite{bit}, ChangeFormer combines a hierarchically structured {Transformer} encoder with a Multi-Layer Perception (MLP) decoder. 
Moreover, the same authors of \cite{changeformer}, in \cite{semicd}, proposed a semi-supervised CD model in which they added an unsupervised loss term to the supervised one. The new term is computed constraining the output change probability map of the unlabeled bitemporal image pair to be consistent, while a small random perturbations is applied on the deep feature difference map.

Lately, several urban mapping studies have been started to integrate EO data fusion into supervised DL CD workflows \cite{nascetti_SAR_optical_2D}.
SAR imagery, indeed, captures unique information about the physical properties of the Earth's surface and is less affected by weather conditions than optical images.
Nevertheless, the uni-modal application of SAR images in supervised multi-temporal CD analyses is limited by the difficulty of collecting pixel-wise labeled SAR imagery \cite{suggested_reviewer1_SAR}. 
Moreover, network architectures which perform well on single sensor data may not be suitable to process multi-sensor data \cite{nascetti_SAR_optical_2D}. Therefore,  advanced network architectures should be considered to integrate optical and SAR images effectively \cite{nascetti_SAR_optical_2D}.
In particular, individual streams based on the U-Net architecture were introduced in \cite{nascetti_SAR_optical_2D} to process SAR and optical images separately, before combining extracted features at a later decision stage. 
\cite{suggested_reviewer1_SAR} proposed, instead, an unsupervised CD approach to train a cycle-consistent GAN (CycleGAN \cite{cycleGAN}) network that was used as a multi-temporal optical-like deep feature extractor from  unlabeled bitemporal very-highspatial-resolution SAR images;  then a deep CVA and fuzzy rules are applied to detect changed buildings.
However, further studies are needed to investigate the use of DL algorithms for SAR and optical integration, before applying them at decision level for urban CD \cite{nascetti_SAR_optical_2D},
but this is beyond the scope of this paper.

\subsubsection{3D Change Detection}

Different kinds of imagery and approaches can be used to monitor 3D changes. 
However, differently from 2D CD applications for which SAR imagery has recently started to be integrated within the proposed pipelines, the few approaches developed so far for solving the 3D CD task are based exclusively on the use of optical images.

{In particular, 3D CD applications can rely on two diverse approaches to detect 3D changes: the DSM difference or the cloud-to-cloud distance \cite{3D_CD_review,RemoteSensing_3d_CD}. 
The first approach computes a per pixel height difference between bitemporal DSMs in the format of a 2.5D map -- this map is what we call 3D CD map in our workflow. Conversely, the second approach directly computes the 3D Euclidean distances between the points of one point cloud and the mesh built on the other point cloud used as reference.
Due to its simplicity, the first approach is still the most used in EO applications, whereas only lately methods using the point clouds based approach have been developed \cite{RemoteSensing_3d_CD, suggested_reviewer1_volumetric_3DCD}.
Indeed, 3D CD maps convey the elevation change information in the same format of 2D CD maps, thus they can be easily integrated in standard 2D CD pipelines, whereas point cloud data are unordered and irregularly distributed, making the extraction of the information they carry more difficult.}

{Specifically}, a critical review of the last developments and applications of 3D CD using RS and close-range data is given in \cite{3D_CD_review}.
More recently, \cite{lidar_CD} carried out a survey of the airborne LiDAR CD methods currently employed in Earth science applications. 
However, this survey did not address urban change detection \cite{RemoteSensing_3d_CD}, one of the most challenging CD scenarios due to the complex nature of urban spaces \cite{nascetti_SAR_optical_2D}.
Finally, \cite{3D_CD_buildings,RemoteSensing_3d_CD,suggested_reviewer1_volumetric_3DCD}, are the only available studies where a comparative analysis for 3D CD was proposed \cite{3DCD_nostro}. In the first study \cite{3D_CD_buildings}, five methods were compared considering two criteria: the ability to detect demolished and new buildings, and the capability to provide information about the magnitude of the elevation changes. However, no quantitative results were provided, and the evaluation was carried out on a private dataset \cite{3D_CD_buildings}. Moreover, this study did not employ any DL algorithms.
On the other hand, in \cite{RemoteSensing_3d_CD}, an original simulator of multi-temporal aerial LiDAR urban point clouds was developed to automatically build a labelled 3D CD dataset consisting of pairs of 3D point clouds annotated according to the synthetic changes imposed by the authors. Six different 3D CD methods were assessed either by directly using the 3D point clouds or employing the DSMs generated from their rasterization \cite{RemoteSensing_3d_CD}. In particular, the authors compared traditional methods, such as the application of different thresholding and filtering algorithms on the input DSMs, with both a machine learning algorithm (a random forest fed with hand-crafted features) and two DL networks (consisting in a feed-forward network and a Siamese network) \cite{RemoteSensing_3d_CD}. 
{
In \cite{suggested_reviewer1_volumetric_3DCD}, a workflow for 3D CD was developed implementing an automated progressive bundle adjustment 
of multitemporal UAV-oblique images for the production of multitemporal point clouds where the areas affected by changes and the corresponding volumes are detected with a coarse-to-fine strategy.}
Nevertheless, all the aforementioned studies \cite{lidar_CD,3D_CD_buildings,RemoteSensing_3d_CD} employed exclusively 3D data as input and no optical imagery {(in \cite{suggested_reviewer1_volumetric_3DCD}, the images are used to build the multitemporal point clouds and not to directly detect the volumetric changes)}, still leaving a strong barrier {to a wide applicability of these methods to the 3D CD task}.

Finally, the authors of \cite{pan2022self} proposed a multi-path self-attentive hybrid coding network model (MAHNet) that fuses high-resolution satellite imagery and DSMs for CD of urban buildings. The model was tested on a private and self-produced dataset made of images captured by the Gaofen-7 stereo mapping satellite \cite{pan2022self}. In this approach, however, the 3D information retrieved from the DSMs was only used to enhance the 2D change detection results, i.e the identification of the planimetric extent of the areas subjected to elevation changes, and no information about the corresponding elevation variation was returned in the output CD maps, making this contribution very {different} from ours.

Considering the aforementioned methods, to our knowledge, there is not yet an algorithm able to solve the 3D CD task starting from bitemporal optical images, neither as a single task or in a multitask fashion. In the {following section}, we {hence} propose a {Transformer}-based network to simultaneously solve the 2D and 3D CD tasks.

\section{Methodology}
\label{sec:meth}
The aim of this work is to develop a 2D-3D CD network able to generate two different CD maps starting from {bitemporal} optical images: a 2D CD map identifying the pixels affected by an elevation change and a 3D CD map quantifying the value of the elevation change for every pixel. 
Specifically, we propose a {Transformer}-based network, the MultiTask Bitemporal Images {Transformer} (MTBIT).
MTBIT extends BIT \cite{bit}, with the aim to solve both the 2D CD and the newly proposed 3D CD task simultaneously. The architecture is composed of four parts: i) a Siamese semantic tokenizer, which generates a set of tokens (i.e. visual words) for each optical input image; ii) a {Transformer}-based encoder, which encodes positional and context information in the token space; iii) a {Transformer}-based decoder, that extracts feature maps projecting the tokens in the pixel space; iv) a pair of prediction heads which allow double mapping from the refined feature maps to 2D and 3D prediction maps. 
Particularly, a Siamese CNN backbone is fed with two input optical images $X_1$ and {$X_2 \in \mathbb{R}^{H {\times} W \times B}$, where H and W are  the height and the width of the input images, and B is the number of the input bands,}  to extract, respectively, two {sets of feature maps} $Z_1$ and {$Z_2 \in \mathbb{R}^{H' {\times} W' \times C}$, where H' is the height and W' is the width and C are the feature maps channels}. To get a vocabulary of semantic tokens (i.e. visual words), these {feature maps} pass through a two-step semantic tokenizer made of:
a)	spatial attention maps, that is a {CNN layer, followed by a} softmax, operated on the height H' and width W' of each {feature map};
b)	a weighted average sum of pixels of the obtained {maps}. Formally:

\begin{equation}
\begin{split}
&\mathbf{T}=\left(\mathbf{A}\right)^{T} \mathbf{X}\\
\end{split}
\end{equation}

\noindent {where $\mathbf{A} = \text{softmax}(\text{conv}(\mathbf{Z})) \in \mathbb{R}^{H{ \times }W \times L}$} are the attention maps, {where L is the length of the vocabulary}.

The obtained two sets of semantic tokens $\mathbf{T}^{i}$ (one for each image) are concatenated {in one set, namely $\mathbf{T}^{*}\in \mathbb{R}^{2L \times C}$}.
Before passing them through the {Transformer} encoder, the tokens are used to learn positional encodings, useful to encode the position of the token in the space. Given the fact that the encoder consists of $N$ layers of multihead self-attention (MSA) and MLP blocks, the positional encodings are also very useful to have shared representations across heads of the tokens.
Particularly, the semantic tokens are the input of the self-attention matrices {(the query $\mathbf{Q} = \mathbf{T}^{*(l-1)}\mathbf{W}^{q}$, the key $\mathbf{K} = \mathbf{T}^{*(l-1)}\mathbf{W}^{k}$ and the value $\mathbf{V} = \mathbf{T}^{*(l-1)}\mathbf{W}^{v}$. where $\mathbf{W}^{q}, \mathbf{W}^{k}, \mathbf{W}^{v} \in  \mathbb{R}^{C \times d}$, where $d$ is the channel dimension and $l$ the layer)}. The single head-attention is defined as in \cite{vaswani2017attention}.
MSA performs multiple independent attention heads in parallel. The outputs are, then, concatenated and projected. MSAs are shaped in order to capture different information in different representation spaces. From a formal point of view,
{
\begin{equation} 
\operatorname{MSA}(\mathbf{T}^{*(l-1)}) = \operatorname{Concat}(\text{Att}_{1}, ..., \text{Att}_{H})\mathbf{W}^{E}
\end{equation}
with
\begin{equation} 
\text{Att}_{h} = \text{softmax}\left(\frac{\mathbf{Q_{h} K_{h}}^T}{\sqrt{d}}\right) \mathbf{V}_{h} + \mathbf{PE}_{h}
\end{equation}\\
where $h = 1, ..., H$, with H denoting the} number of attention heads, PE are the learnable positional encodings of the encoder and $\mathbf{W}^{E} \in \mathbb{R}^{hd \times C}$ is the weight matrix.
On the other hand, the MLP block is made of two linear projection layers. A Gaussian error linear unit (GELU) module \cite{hendrycks2016gaussian} is inserted after the first linear transformation.

{In this way, we obtained a new set of tokens $\mathbf{S}$ that encodes the context information from the input images. We split it into two set $\mathbf{S}^i$ with $i = 1, 2$ and we pass them through the decoder. The aim of the {Transformer} decoder is to project back to the pixel space the semantic information capable to correctly predict 2D and 3D changes. Practically, the {Transformer} decoder consists of $M$ layers of multihead cross attention (MA) and MLP blocks. In MA, $\mathbf{Q} = \mathbf{Z_i}^{(l-1)}\mathbf{W}^{q}$, $\mathbf{K} = \mathbf{S}^{(l-1)}\mathbf{W}^{k}$ and $\mathbf{V} = \mathbf{S}^{(l-1)}\mathbf{W}^{v}$.}

Finally, the output feature maps from the decoder are passed through the pair of prediction heads, each consisting in a Fully Convolutional Neural Network (FCN). {Particularly, given two feature maps $\mathbf{Y}_{1}$ and $\mathbf{Y}_{2} \in \mathbb{R}^{H''\times W'' \times C}$, the final predictions maps, $\mathbf{M}_{2D} \in \mathbb{R}^{H \times W \times 2}$ and $\mathbf{M}_{3D} \in \mathbb{R}^{H \times W}$, are obtained in the following way:}

\begin{equation}
\label{eq:pred_head_2d}
\mathbf{M}_{2D}=\text{sigmoid}\left(f_1\left(\mathbf{Y}_{2}-\mathbf{Y}_{1}\right)\right)
\end{equation}
{and}
\begin{equation}
\label{eq:pred_head_3d}
\mathbf{M}_{3D}=\tanh\left(f_2\left(\mathbf{Y}_{2}-\mathbf{Y}_{1}\right)\right)
\end{equation}

{where $f_{1}$ and $f_{2}$ are 3x3 convolutional layers.}

{As loss of the proposed network, we use a weighted combination of a 2D loss, taking as input $\mathbf{M}_{2D}$, and a 3D loss, taking as input   $\mathbf{M}_{3D}$:}
\begin{equation}
\label{eq:loss}
    \text {L}= \alpha \cdot \text {L}_{2D} + \beta \cdot \text {L}_{3D}
\end{equation}
where $\alpha$ and $\beta$ are fixed weights.
In particular, we use a binary cross-entropy (BCE) for ${L}_{2D}$ and the mean squared error (MSE) for ${L}_{3D}$.

\begin{figure*}
  \centering
  \includegraphics[width=1\linewidth]{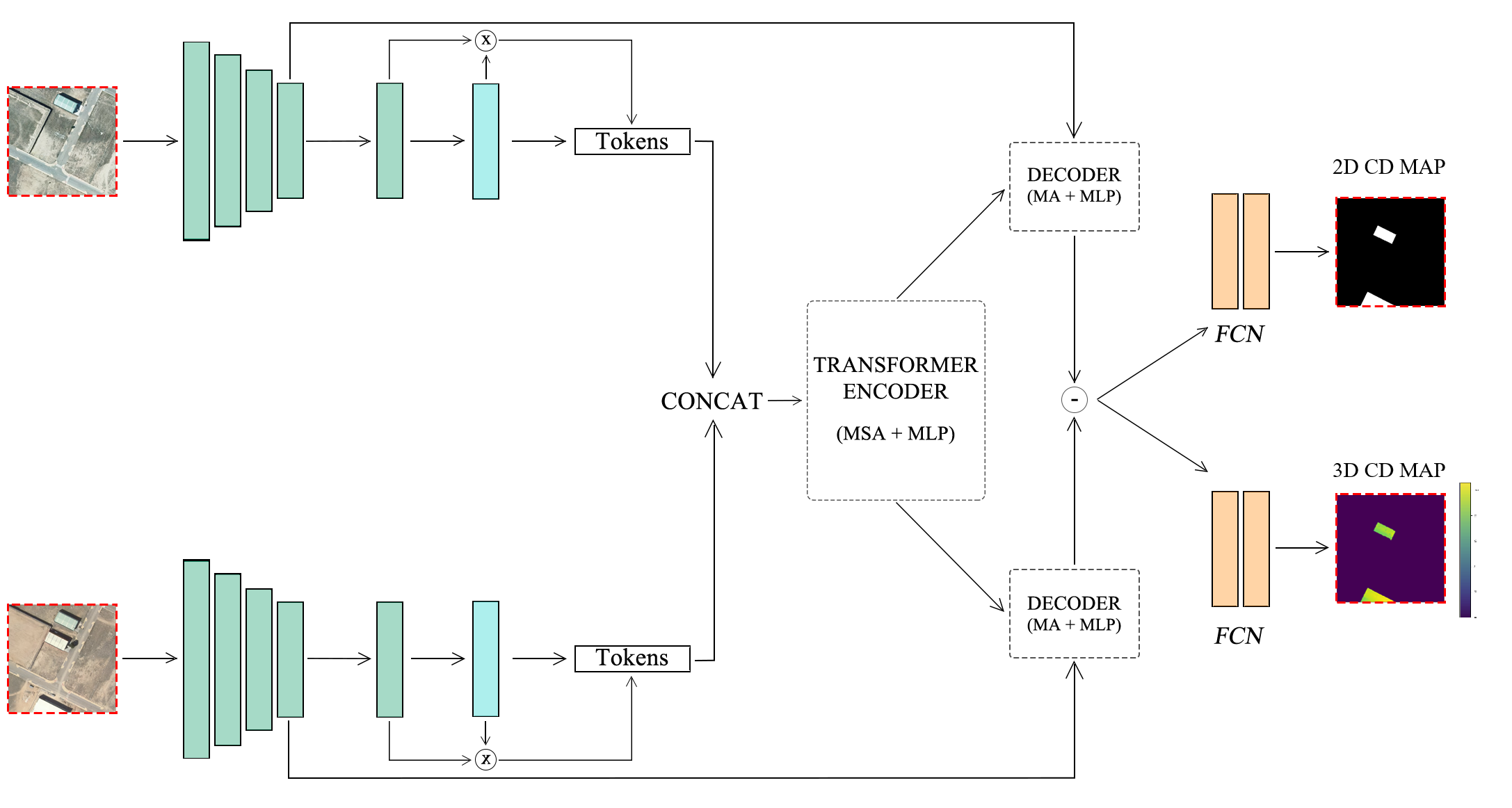}
  \caption{Schematic representation of the MultiTask Bitemporal Image {Transformer} (MTBIT) network architecture.}
  \label{fig:arch_mtbit}
\end{figure*}

The MTBIT architecture is summarized in Figure \ref{fig:arch_mtbit}.

\section{Dataset}
\label{sec:dataset}

The dataset used to evaluate the potentialities of the proposed 2D-3D CD network is a {modified} version of the previously presented 3DCD dataset \cite{3DCD_nostro}, freely available at \url{https://sites.google.com/uniroma1.it/3dchangedetection/home-page}. 
The dataset, which can be classified as a top-view dataset according to the taxonomy of visualization scenarios proposed in \cite{3D_CD_review}, is composed of:
\begin{itemize}
    \item 472 pairs of images cropped from optical orthophotos acquired through two different aerial surveys (performed respectively in 2010 and in 2017);
    {\item 472 pairs of DSMs produced from the rasterization of point clouds acquired through two different LiDAR flights covering the same area and years (2010 and 2017) of the orthophotos\footnote{{The DSMs were not released in the previous version of the dataset.}};}
    \item the corresponding 472 2D CD maps in raster format;
    \item the corresponding 472 3D CD maps -- the elevation variation ($\Delta H${$=H_{2017}-H_{2010} $}) maps -- in raster format.
    \end{itemize}
    
The original orthophotos and the LiDAR point clouds used to produce the dataset were made available by the \textit{Centro de Descarga Nacional} \cite{CNIG}.
{The {modified} version of the dataset includes also the georeferencing information: the reference system is EPSG:3042 - ETRS89 / UTM zone 30N (N-E), i.e. the reference system of the original orthophotos and LiDAR point clouds.}

In particular, the dataset covers the historical and urban centre of the city of Valladolid in Spain, including {also} the surrounding commercial areas. 
Agricultural zones were not considered since no significant elevation changes were observed in these areas. Furthermore, only the changes affecting artificial structures, such as the construction and the demolition of buildings/roads/bridges, were included in the 2D and 3D CD maps.

\begin{figure*}[h!]
    \includegraphics[width=1\textwidth]{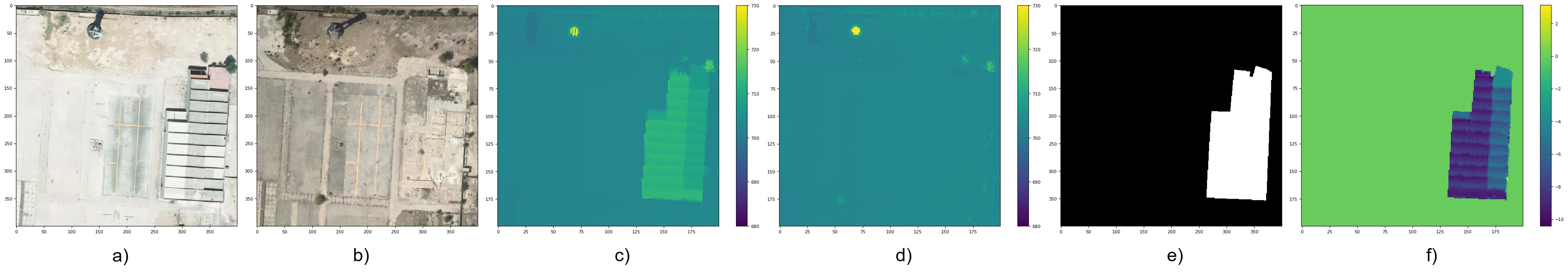}
    \includegraphics[width=1\textwidth]{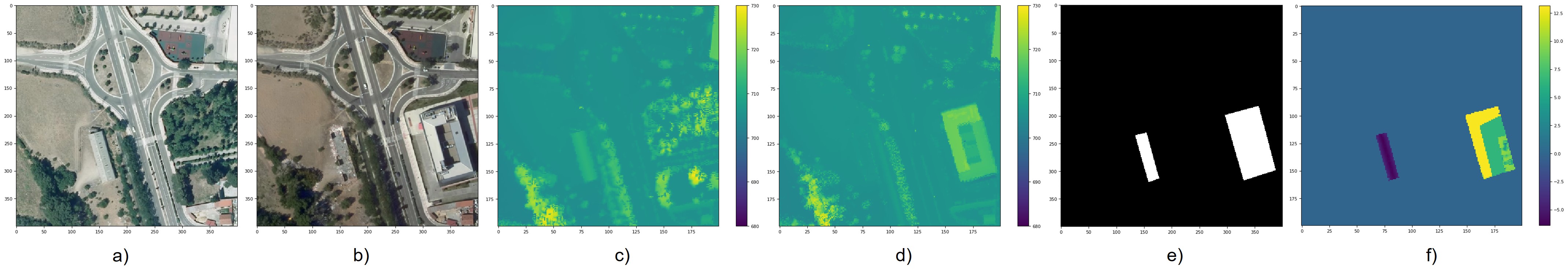}
    \caption{Two examples from the {modified} 3DCD dataset. a) 2010 optical image, b) 2017 optical image, {c) 2010 DSM (the colour bar is expressed in meters), d) 2017 DSM (the colour bar is expressed in meters)}, e) 2D CD map (the pixels affected by a change in elevation are reported in white), f) 3D CD map (the colour bar is expressed in meters): first row: demolition ($\Delta H<0$), second row: construction ($\Delta H>0$).}
    \label{fig:full_ex}
\end{figure*}

Two examples of the 3DCD dataset are shown in Figure \ref{fig:full_ex}. Each tile of the dataset {(including one pair of images, one pair of DSMs, and the corresponding 2D and 3D CD maps)} covers an area equal to 200 m $\times$ 200 m. 

The optical images (Figure \ref{fig:full_ex}-a and \ref{fig:full_ex}-b) and the 2D CD maps (Figure \ref{fig:full_ex}-e), which we produced from the orthophotos, have a size of 400×400 pixels with a GSD of 0.5 m.

The 2D CD maps are binary maps, in which we annotated the pixels belonging to the areas affected by a change in elevation. Specifically, the 2D CD maps include the following two classes (Figure \ref{fig:full_ex}-e): (i) no change (null elevation change in the corresponding 3D CD map: $\Delta H=0$); (ii) changes due to construction (positive elevation change in the corresponding 3D CD map: $\Delta H>0$) or demolition (negative elevation change in the corresponding 3D CD map: $\Delta H<0$) of anthropic artefacts (e.g. buildings, roads, bridges). 

\begin{figure}[h]
    \centering
    \begin{subfigure}[h]{0.32\textwidth}
      \centering
          \includegraphics[width=1\textwidth]{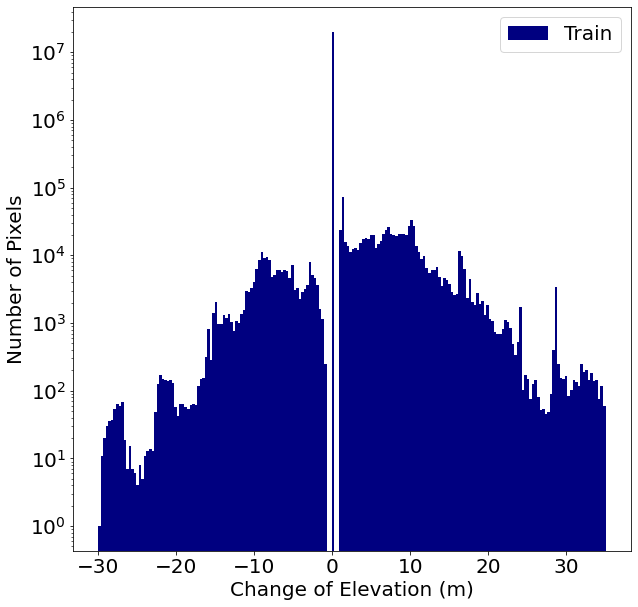}
    \caption{Pixel distribution in train}
    \label{fig:Train}
    \end{subfigure}
    \hfill
    \begin{subfigure}[h]{0.32\textwidth}
      \centering
          \includegraphics[width=1\textwidth]{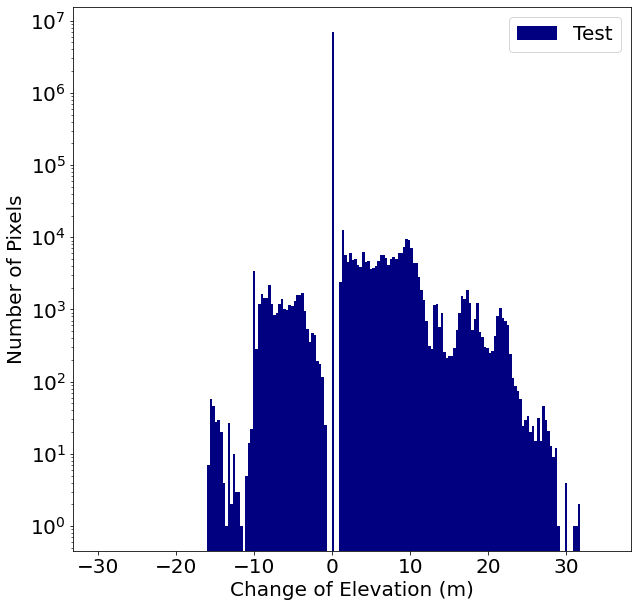}
    \caption{Pixel distribution in test}
    \label{fig:Test}
      \end{subfigure}
    \begin{subfigure}[h]{0.32\textwidth}
      \centering
          \includegraphics[width=1\textwidth]{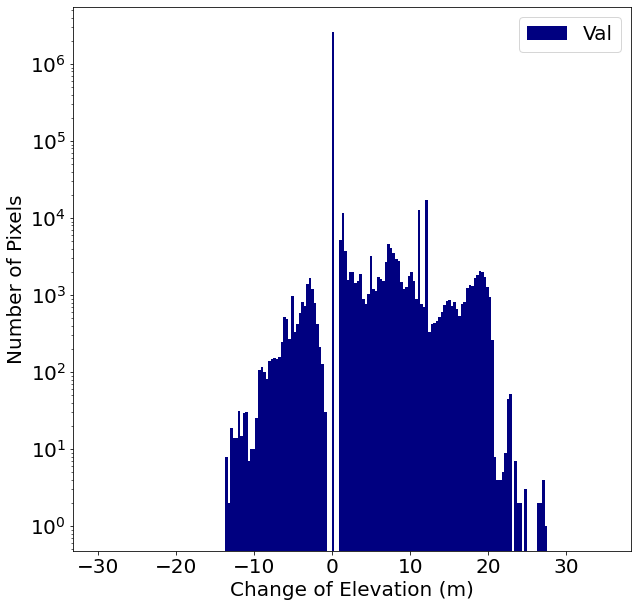}
    \caption{Pixel distribution in val}
    \label{fig:Val}
      \end{subfigure}
     \caption{Distribution of the elevation change values among all the pixels of the 3D CD maps in each subset of the {modified version of the} 3DCD dataset. \label{fig:distributions_dH}}
\end{figure}

The 3D CD maps (Figure \ref{fig:full_ex}-f) have a size of 200×200 pixels with a GSD of 1 m{, as well as the DSMs (Figure \ref{fig:full_ex}-c and \ref{fig:full_ex}-d), }and were obtained from the per-pixel differences between {the} two different LiDAR DSMs ($\Delta H = H_{2017} - H_{2010}$). In particular, through a visual comparison with the corresponding orthophoto pair, we only considered pixels affected by a real elevation change, while the $\Delta H$ of the pixels containing only noise was set to zero. In the 3D CD maps, the values of the elevation changes are represented on a continuous range varying from -30 m to {-1 m and from 1 m to} 35 m {(Figure \ref{fig:distributions_dH})}. Indeed, all the elevation changes lower than one meter in absolute value were considered as negligible with respect to the entity of elevation variations usually affecting man made artefacts, and for this reason their value was set to zero \cite{3DCD_nostro}. 

The main features of the data contained in the 3DCD dataset are recalled in Table \ref{tab:features}.

Finally, the dataset was split into \textit{train}, \textit{validation} and \textit{test} subsets to allow an easy and reproducible benchmarking.
In particular, the train subset contains 320 images ($\sim68\%$), the test subset 110 images ($\sim23\%$) and the validation subset 42 images ($\sim9\%$). The division was performed maintaining a similar distribution of changed pixels (2D CD maps) and of elevation changes (3D CD maps) among the three subsets (see Table \ref{Per} and Figures \ref{fig:Train}, \ref{fig:Val}, \ref{fig:Test} for details){, verifying that the subsets did not share the same buildings in the {modified} version of the dataset, removing hence any spatial autocorrelation.}

\begin{table}[h]
\centering
\caption{Main features of the data contained in the {modified version of the 3DCD} dataset.}\label{tab:features}
\begin{tabular}{|c|c|c|}
\hline
\textbf{Image} & \textbf{Number of pixels} & \textbf{GSD}\\
\hline
\textit{Orthophotos from 2010} & 400$\times$400 & 0.50 m\\
\hline
\textit{Orthophotos from 2017} & 400$\times$400 & 0.50 m\\
\hline
\textit{DSMs from 2010} & 200$\times$200 & 1.0 m\\
\hline
\textit{DSMs from 2017} & 200$\times$200 & 1.0 m\\
\hline
\textit{2D CD maps} & 400$\times$400 & 0.50 m\\
\hline
\textit{3D CD maps} & 200$\times$200 & 1.0 m\\
\hline
\end{tabular}
\end{table}

\begin{table}[h]
\centering
\caption{Percentage of pixels characterised by `change' and `no change' for each subset.}\label{Per}
\begin{tabular}{|c|c|c|c|}
\hline
\textbf{Subset} & \textbf{Number of Images} &  \textbf{Change} & \textbf{No change}\\
\hline
\textit{Train} & 320 & 5\% & 95\%\\
\hline
\textit{Test} & 110 & 4\% & 96\%\\
\hline
\textit{Val} & 42 & 5\% & 95\%\\
\hline
\end{tabular}
\end{table}

\section{Experimental setup}
\label{sec:setup}

For the training phase, a single Tesla T4 16 GB GPU was used. 
We trained the proposed network -- MTBIT -- for 300 epochs with the AdamW optimizer and a learning rate of 0.0001. Moreover, we used a fixed batch size of 15 and we performed the following augmentation strategies on the input images: resizing to $256\times256$, random horizontal flip, random geometric transformation (i.e. shifting, scaling and rotating), addition of random gaussian noise, random radiometric transformation (i.e. brightness, sharpening, blurring, contrasting, saturation). Furthermore, we normalized the 3D CD maps between -1 and 1, using a TanH as last non-linear layer of the proposed network. For the 2D loss (i.e. BCE), we set the weights to 0.05 for no-change pixels and 0.95 for change ones. 
In addition, we set $\alpha = 1$  and $\beta = 3$ in Equation \eqref{eq:loss} {(see Section \ref{sec:loss_w} for further details)}.

To evaluate the performance of the proposed network, we opted for two 2D metrics, i.e. intersection over union (IoU) and F1-score (F1), {referred to the change class (that is 1)}, and two 3D metrics, i.e. root mean squared error (RMSE) and changed root mean squared error (cRMSE). Particularly, cRMSE is a RMSE that considers only the error with respect to the ground truth pixels affected by an actual elevation change (i.e. $\Delta H\neq 0$). Formally, the metrics are expressed as follows:
\begin{equation}
\text{IoU} = \frac{\text{TP}}{\text{TP} + \text{FN} + \text{FP}}\\
\end{equation}

\begin{equation}
\text{F1} = \frac{2\text{TP}}{2\text{TP}+\text{FP}+\text{FN}}
\end{equation}

\begin{equation}
\text{RMSE} = \sqrt{\frac{1}{n}\sum_{i=1}^{n}\left(\widehat{\Delta H}_{i} - \Delta H_{i}\right)^{2}}
\end{equation}

\begin{equation}
\text{cRMSE} = \sqrt{\frac{1}{n_c}\sum_{i=1}^{n_c}\left(\widehat{\Delta H}_{i}^{C} - \Delta H_{i}^{C}\right)^{2}}
\end{equation}

\noindent where TP stands for True Positives, TN for True Negatives, FP for False Positive and FN for False Negatives, $\widehat{\Delta H}_{i}$ for predicted pixels, $\Delta H_{i}$ for ground truth pixels, $\widehat{\Delta H}_{i}^{C}$ for predicted pixels, $\Delta H_{i}^{C}$ for respective ground truth changed pixels. Between the 3D CD metrics, cRMSE is preferable to standard RMSE because the latter is less sensible to changes (and errors), being most of the changes in the target map at 0 value and thus easy to be detected.
In addition to several ablation studies to gain a better understanding of the behaviour of the proposed architecture, we performed a comparison with different existing networks designed for the 2D CD, given the fact that, to our knowledge, no architecture able to solve the 3D CD task is currently available. Specifically, to make these existing networks able to predict the 3D CD maps, we changed the prediction head with a pair of FCNs, able to solve both the 2D and 3D CD tasks. We selected the following models for the comparison: ResNet18, ChangeFormer \cite{changeformer}, FC-EF \cite{daudt2018fully}, FC-Siam-diff \cite{daudt2018fully} and a modified version of FC-Siam-conc, that we named SUNet (Siamese ResU-Net).

Particularly, SUNet is made of: a) a Siamese encoder; b) a convolutional decoder; c) a pair of prediction heads. The Siamese encoder consists of two ResNet architectures with shared weights, each of which extracts the features from each input image. The choice of the ResNet is pivotal for the effective learning of the features, making possible to avoid the degradation problem. At the end of the ResNet encoder, the features are concatenated to be fed through a convolutional decoder, equipped with skip connections. Specifically, we opted for concatenating the skip connection of each layer of the encoder. This choice is justified by the structure of our problem, for which not all the land use changes corresponding to different features match a 2D change in the prediction maps. Finally the decoded feature maps are passed through a prediction head pair, that, as for the MTBIT, is made of two FCNs. The code is available at \url{https://sites.google.com/uniroma1.it/3dchangedetection/home-page}.

{Moreover, we also provide a baseline obtained using IM2HEIGHT \cite{mou2018im2height}, a network designed to infer an elevation map from a single RGB image. We trained this model twice, firstly on the 2010 images and then on the 2017 images. After inferring the resulting elevation maps on the test images, we subtracted the 2010 elevation maps to the 2017 elevation maps, obtaining the baseline 3D CD maps. For this experiment, we started from the code available at \url{https://github.com/dettmar/im2height}.}

\section{Experimental Results}
\label{sec:exp_res}

As described in Section \ref{sec:meth}, several models were trained with the final goal of automatically producing both the 2D and 3D CD maps, using a pair of bitemporal optical images as input.
In the following sections, the best results and different ablation studies are described.

\subsection{Comparison}\label{sec:comparison}

\begin{figure}[p]
    \centering
    \begin{subfigure}[h]{0.48\textwidth}
      \centering
          \includegraphics[width=1.1\textwidth]{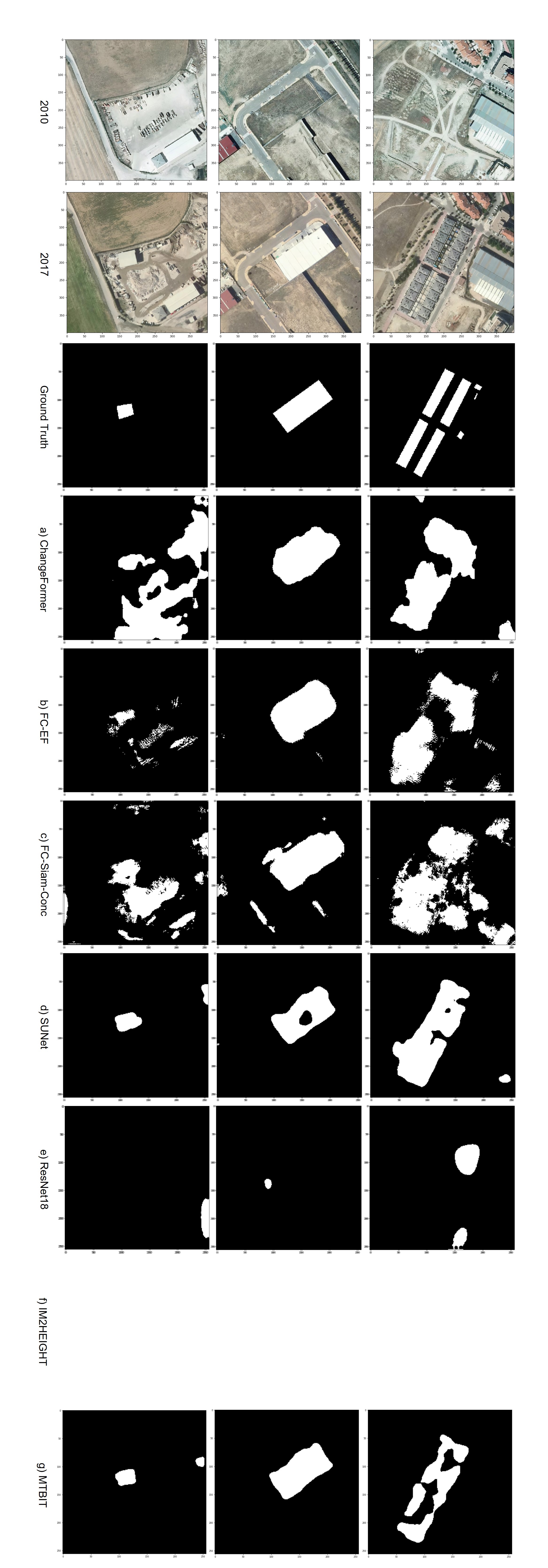}
    \caption{2D CD}
    \label{fig:2d}
    \end{subfigure}
    \hfill
    \begin{subfigure}[h]{0.48\textwidth}
      \centering
          \includegraphics[width=1.1\textwidth]{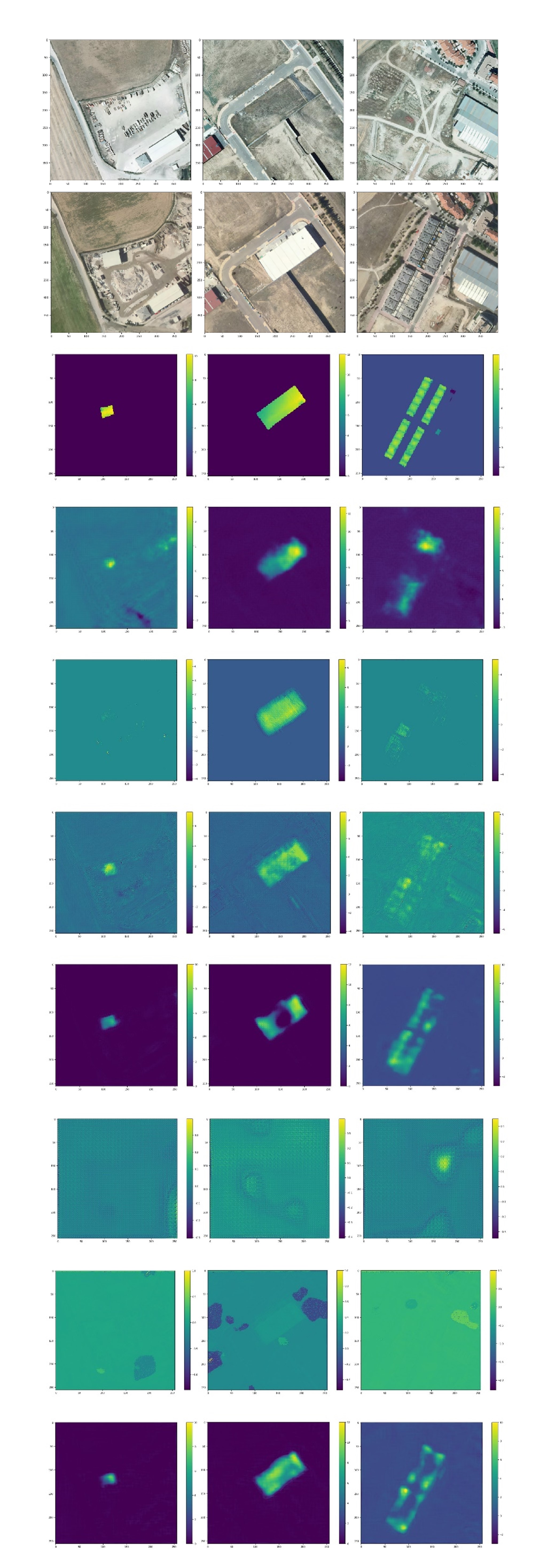}
    \caption{3D CD}
    \label{fig:3d}
      \end{subfigure}
\caption{Visualisation of 2D CD and 3D CD maps for the models described in Table \ref{tab:comp}.}
\label{fig:comp}
\end{figure}

\begin{figure}[h!]
    \centering
    \begin{subfigure}[h]{0.32\textwidth}
      \centering
          \includegraphics[width=1\textwidth]{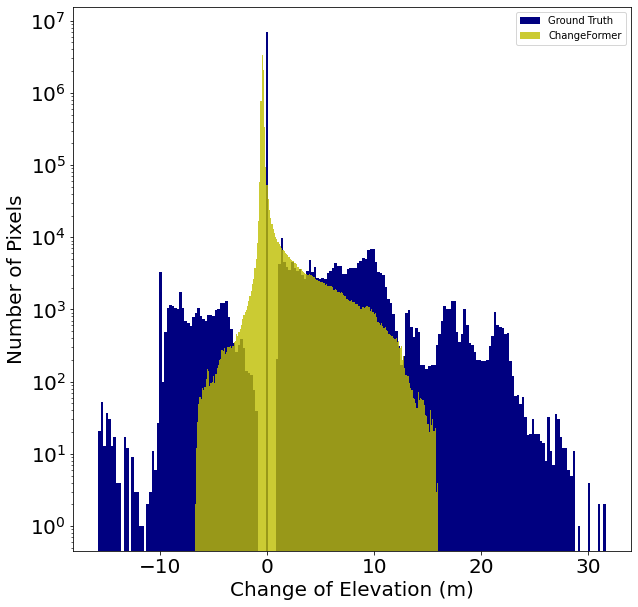}
    \caption{Change Former}
    \label{fig:cf}
    \end{subfigure}
    \hfill
    \begin{subfigure}[h]{0.32\textwidth}
      \centering
          \includegraphics[width=1\textwidth]{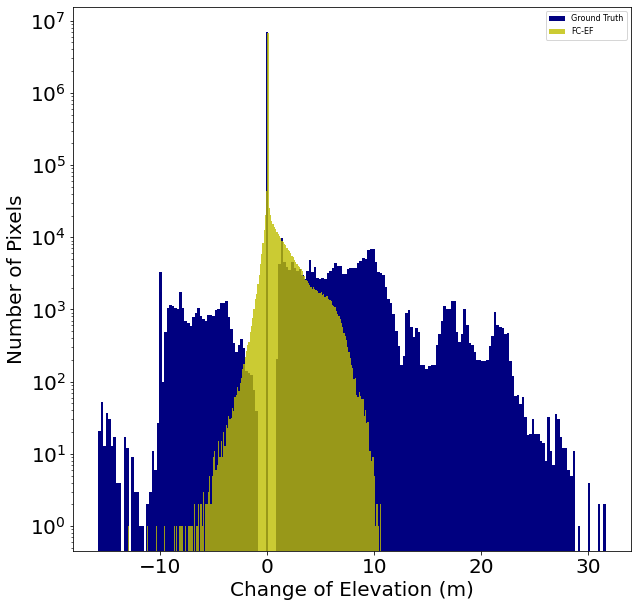}
    \caption{FC-EF\\ }
    \label{fig:unet}
      \end{subfigure}
    \begin{subfigure}[h]{0.32\textwidth}
      \centering
          \includegraphics[width=1\textwidth]{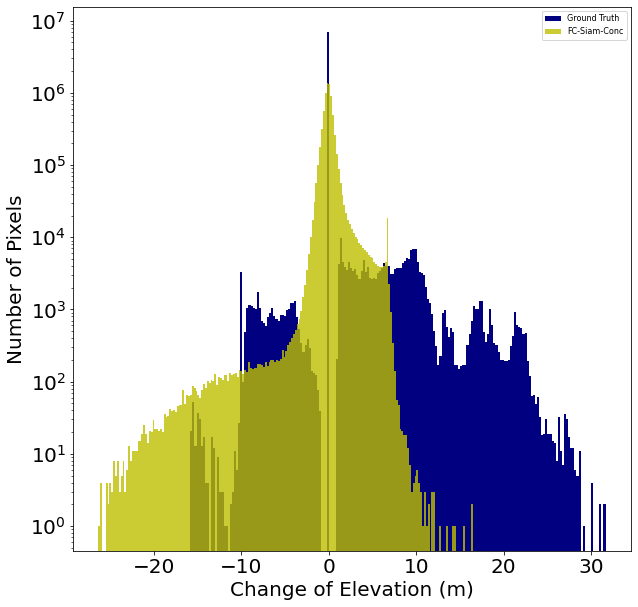}
    \caption{FC-Siam-conc\\ }
    \label{fig:su}
      \end{subfigure}
\medskip
\begin{subfigure}[h]{0.32\textwidth}
      \centering
    \includegraphics[width=1\textwidth]{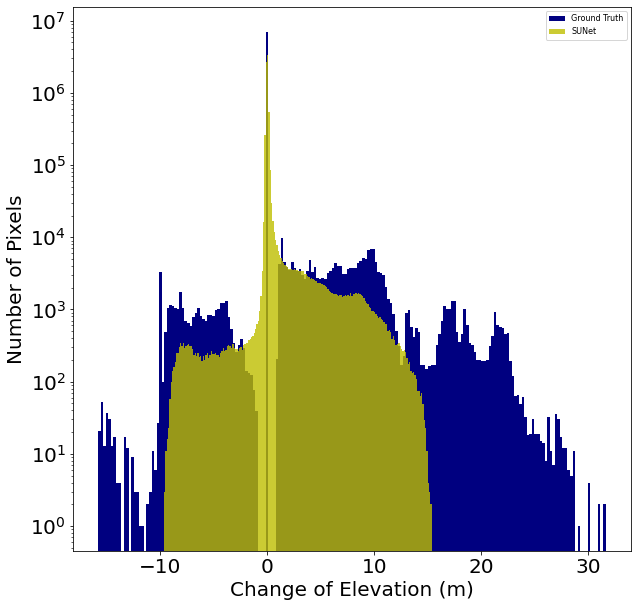}
    \caption{SUNet}
    \label{fig:sunet}
    \end{subfigure}
    \hfill
    \begin{subfigure}[h]{0.32\textwidth}
      \centering
    \includegraphics[width=1\textwidth]{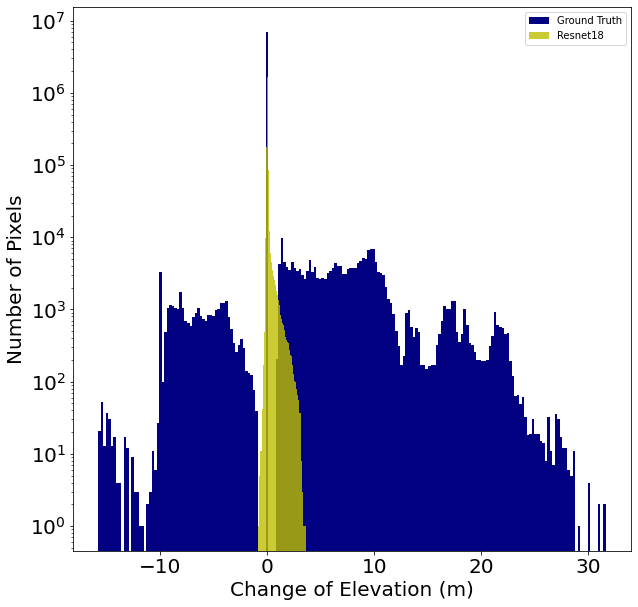}
    \caption{Resnet18}
    \label{fig:resnet}
      \end{subfigure}
    \begin{subfigure}[h]{0.32\textwidth}
      \centering
    \includegraphics[width=1\textwidth]{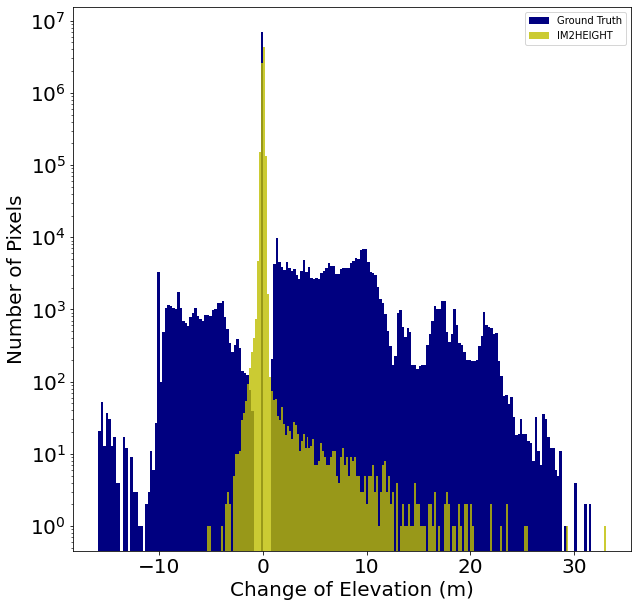}
    \caption{IM2HEIGHT}
    \label{fig:im}
      \end{subfigure}
    \begin{subfigure}[h]{0.32\textwidth}
      \centering
    \includegraphics[width=1\textwidth]{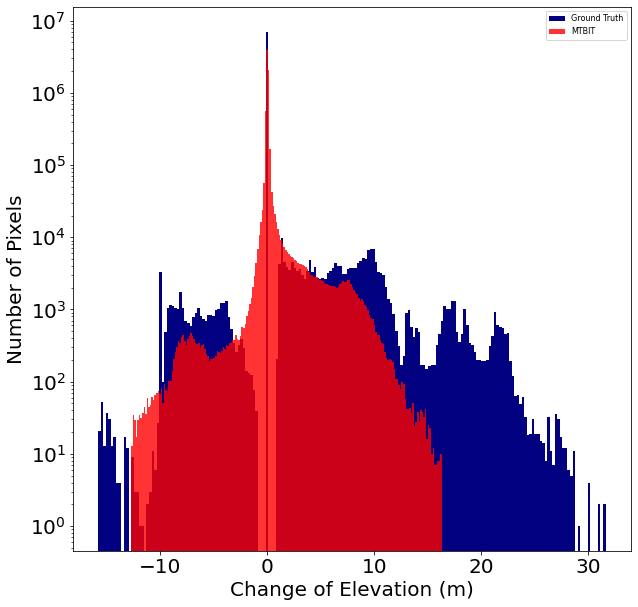}
    \caption{MTIBIT}
    \label{fig:bit}
      \end{subfigure}
\caption{Comparison between the distribution of the elevation change values in the ground truth (in blue) and the distribution of the predicted elevation change values obtained with the different models reported in Table \ref{tab:comp}.}
\label{fig:isto}
\end{figure}

We compared the proposed network, MTBIT, with the aforementioned architectures to evaluate its performance.
{In particular}, in addition to the quantitative {assessment of the results, performed evaluating the metrics described in} Section \ref{sec:setup} {(Table \ref{tab:comp})}, we performed also two qualitative comparisons: we visually compared the {predicted} 2D and 3D {CD} maps with the corresponding ground truth {map}s {(Figure \ref{fig:comp})} and we compared the distribution of the predicted elevation changes with the corresponding distribution of the ground truth values (Figure \ref{fig:isto}).

\begin{table}[h!]
\caption{Validation results of the different networks applied to the {modified version of the} 3DCD dataset.}
\label{tab:comp}
\resizebox*{1\textwidth}{!}{\begin{tabular}{|c|c|c|cc|cc|}
\hline
\multirow{2}{*}{Architecture} & \multirow{2}{*}{Figure \ref{fig:comp}}&  \multicolumn{1}{l|}{\multirow{2}{*}{\begin{tabular}[c]{@{}l@{}}Trainable\\ parameters\\      (M)\end{tabular}}} & \multicolumn{2}{c|}{\begin{tabular}[c]{@{}c@{}}2D CD\\ metrics\end{tabular}} & \multicolumn{2}{c|}{\begin{tabular}[c]{@{}c@{}}3D CD\\ metrics\end{tabular}} \\ \cline{4-7} 
 && \multicolumn{1}{l|}{} & \multicolumn{1}{c|}{\begin{tabular}[c]{@{}c@{}}F1\\ (\%)\end{tabular}} & \begin{tabular}[c]{@{}c@{}}IoU\\   (\%)\end{tabular} & \multicolumn{1}{c|}{\begin{tabular}[c]{@{}c@{}}RMSE\\   (m)\end{tabular}} & \begin{tabular}[c]{@{}c@{}}cRMSE\\   (m)\end{tabular} \\ \hline
 ChangeFormer & a & 1,8 & \multicolumn{1}{c|}{42,89} & 27,30 & \multicolumn{1}{c|}{1,31} & 7,09 \\ \hline
 FC-EF& b & 1,3 & \multicolumn{1}{c|}{46,32} & 30,14 & \multicolumn{1}{c|}{1,41} & 7,04 \\ \hline
 FC-Siam-conc & c & 1,5 & \multicolumn{1}{c|}{38,30} & 23,69 & \multicolumn{1}{c|}{1,42} & 10,66 \\ \hline
 SUNet & d &  35,7 & \multicolumn{1}{c|}{59,72} & 42,57 & \multicolumn{1}{c|}{1,24} & 6,47 \\ \hline
 ResNet18 & e & 32,6 & \multicolumn{1}{c|}{15,52} & 8,41 & \multicolumn{1}{c|}{1,56} & 7,45 \\ \hline
 IM2HEIGHT & f & 7,4 & \multicolumn{1}{c|}{-} & - & \multicolumn{1}{c|}{1,57} & 7,59 \\ \hline
\textbf{ MTBIT (ours)} & g & \textbf{13,1} & \multicolumn{1}{c|}{\textbf{62,15}} & \textbf{45,09} & \multicolumn{1}{c|}{\textbf{1,20}} & \textbf{6,46} \\ \hline
\end{tabular}}
\end{table}

Table \ref{tab:comp} shows that the {ResNet18}, ChangeFormer, FC-EF and FC-Siam-conc models provide {the} worst results when applied to the 3DCD dataset, for {both} the 2D {and} the 3D CD task{s}. In fact, the F1 score reaches a value less than 50\% for any of these {four} models and the IoU value is always less than 35\%. Looking at the Figures \ref{fig:2d}-a, \ref{fig:2d}-b and \ref{fig:2d}-c, relative to the 2D CD results, it is evident that the 2D CD is not particularly accurate, both with respect to the detected areas and to their extent. The worst results are obtained with {the ResNet18 architecture which cannot reconstruct correctly the 2D CD maps (Figure \ref{fig:2d}-e). This problem, true also for the 3D CD task, is due to the absence of the skip connections, making the models unsuitable for the proposed tasks (Figure \ref{fig:3d}-e). When adding skip connections and making the network deeper, the results are more accurate. This is the case of SUNet, which reaches results comparable to MTBIT. However, the number of its parameters (35,7 M) is higher than the counterparts. MTBIT, on the other hand, provides the best results for the 2D CD task, obtaining an F1 score of 62.15\% and an IoU value of 45.09\% (Table \ref{tab:comp}-g). From Figure \ref{fig:comp}-g, it can also be observed that changes are detected with high accuracy and the background noise is minimal.}

Also for the 3D CD task, {ResNet18,} ChangeFormer, FC-EF and FC-Siam-conc networks do not achieve accurate results: all the {four} architectures have a cRMSE higher than 7 m, up to more than 10 m for the FC-Siam-conc model (Table \ref{tab:comp}). From the Figures \ref{fig:3d}-a, \ref{fig:3d}-b and \ref{fig:3d}-c, we can see that these models in general show limitations in detecting real changes and have a tendency to underestimate elevation changes. In addition, we can observe that the 3D CD maps predicted by ChangeFormer (Figure \ref{fig:3d}-a) and FC-Siam-conc (Figure \ref{fig:3d}-c) are characterized by a background noise. The latter problem is highlighted by the histograms reported in Figures \ref{fig:cf},  \ref{fig:unet} and \ref{fig:su}; for all the three models, the  distributions of the values of the predicted elevation changes are very different from the distribution of the Ground Truth. The reasons behind these failures vary depending on the considered network: ChangeFormer {-- a fully {Transformer} architecture, designed for 2D CD, which operates on few data -- shows poor generalization capabilities}, detecting changes in LULC that do not correspond to changes in elevation; {conversely}, FC-EF and FC-Siam-Conc are not able of learning effective features for the solution of the 2D-3D multitask (e.g., low number of parameters, lack of effective encoder).
{Furthermore}, also the IM2HEIGHT-based baseline shows a poor performance (Table \ref{tab:comp}-f and Figure \ref{fig:comp}-f). This behaviour is mainly due to the fact that IM2HEIGHT is focused on reconstructing the correct height of the  {single} DSMs, and not on estimating the elevation changes occurred between two epochs (2010 and 2017). Moreover, any small error in reconstructing the DSMs for each of the two epochs is propagated through their difference.

As far as the 3D CD task is concerned, MTBIT achieves the best results with a cRMSE of 6.46 m (Table \ref{tab:comp}-g), even if the 3D CD maps are characterised by a slight background noise (Figure \ref{fig:3d}-g). 
Furthermore, the predicted elevation changes (see Figure \ref{fig:3d}-g) slightly underestimate the real ones. This phenomenon is more evident when analysing the histograms of the elevation changes: the predicted elevation changes follow the ground truth distribution well, but the network fails to reproduce the extreme values with a deviation of about 10 m (Figure \ref{fig:bit}).
Also SUNet provides optimal results for the 3D task, reaching a cRMSE value equal to 6.47 m (Table \ref{tab:comp}-e).

Focusing on the comparison among these two best models (SUNet and MTBIT),
the 2D CD maps obtained using SUNet (Figure \ref{fig:2d}-e) detect changes very well, but are slightly noisier than the 2D CD maps obtained with MTBIT (Figure \ref{fig:2d}-d).
The 3D CD maps predicted using the SUNet model are accurate: the areas are well detected, there is no background noise and the predicted elevation changes are in line with ground truth values (Figure \ref{fig:3d}-d). 
{Moreover, the elevation changes predicted by SUNet follow the distribution of the ground truth values very well (Figure \ref{fig:sunet}).
However}, also {SUNet, like MTBIT,} is not able to predict the extreme values {of the elevation changes}, but with a {high}er deviation than {MTBIT} (Figure \ref{fig:sunet}). 
The problem with the detection of extreme values is probably due to a deficiency in the {training} data: considering that the y-axis of the histograms is logarithmic, the values that the models cannot reproduce are present in less than 100 pixels. 
Also the {3D CD} metrics of MTBIT (RMSE: 1.20 m and cRMSE 6.46 m) and SUNet (RMSE: 1.24 m and cRMSE: 6.47 m) are very similar. Nevertheless, MTBIT has a slightly better ability to identify areas affected by 3D changes, {reaching lower} RMSE and cRMSE values. 
SUNet instead does not perfectly identify the contours of the changed areas, increasing the RMSE, still providing however elevation change values very consistent with the Ground Truth.

{In conclusion}, MTBIT is more appropriate than SUNet for its lower number of parameters. 
Indeed, MTBIT obtains results comparable to the ones obtained by SUNet with a third of the parameters (13,1 M vs 35,7 M). 
This is even more true if we compare the number of parameters of these two models with the other ones used for comparison. In fact, except for ResNet18 (32,6 M), all the other models have about 1.5 M of parameters. IM2HEIGHT, designed for inferring height from RGB images, has 7,4 M of parameters.

In the next sections, we describe in more detail {the} different ablation studies related to the proposed model.

\subsection{Ablation Studies}
\label{sec:abl}

\subsubsection{Components of MTBIT}
\label{sec:bit_abl}

\begin{figure}[p]
    \centering
    \begin{subfigure}[h]{0.48\textwidth}
      \centering
          \includegraphics[width=1\textwidth]{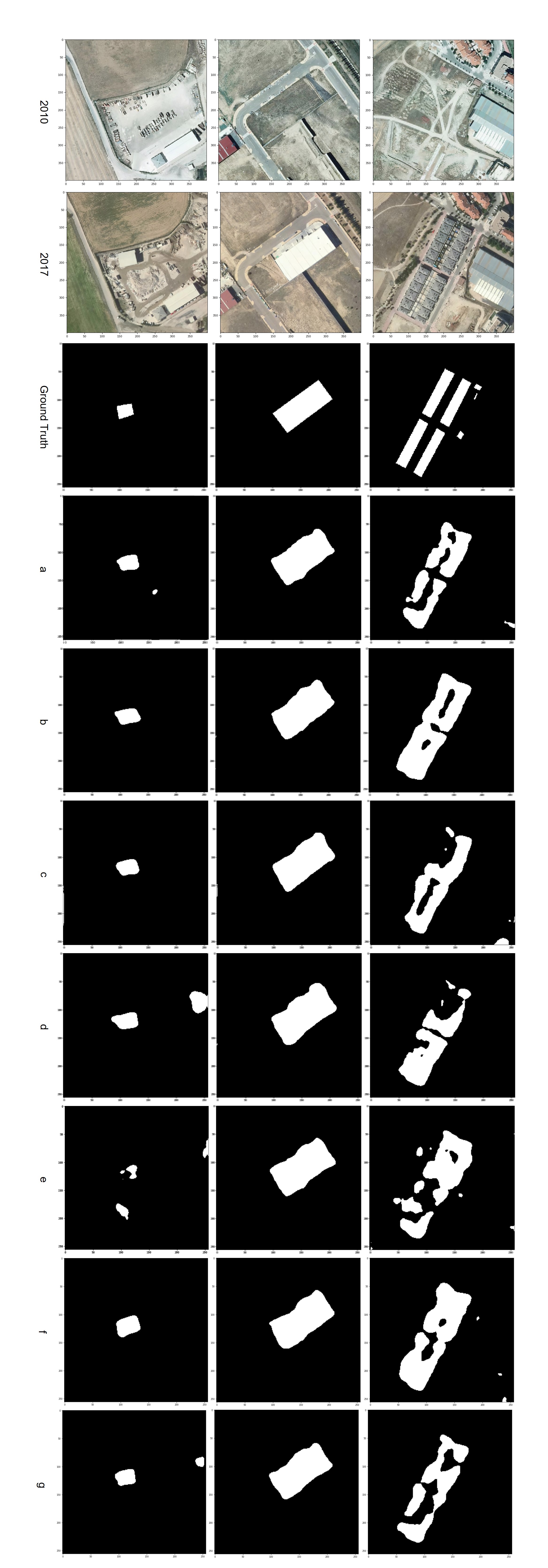}
    \caption{MTBIT 2D CD}
    \label{fig:bit2d}
    \end{subfigure}
    \hfill
    \begin{subfigure}[h]{0.48\textwidth}
      \centering
          \includegraphics[width=1\textwidth]{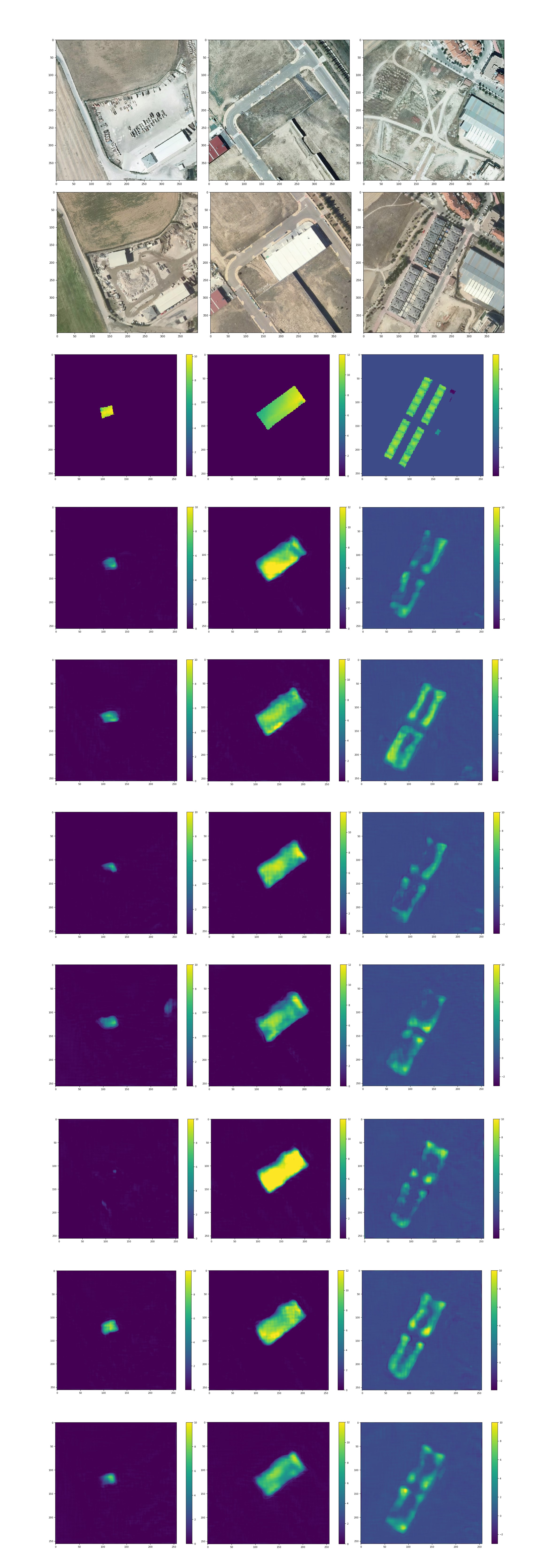}
    \caption{MTBIT 3D CD}
    \label{fig:bit3d}
      \end{subfigure}
\caption{Visualisation of 2D CD and 3D CD maps {obtained with the different} models described in Table \ref{tab:abl1}}
\label{fig:bit_1}
\end{figure}

\begin{figure}[p]
    \centering
    \begin{subfigure}[h]{0.48\textwidth}
      \centering
          \includegraphics[width=1\textwidth]{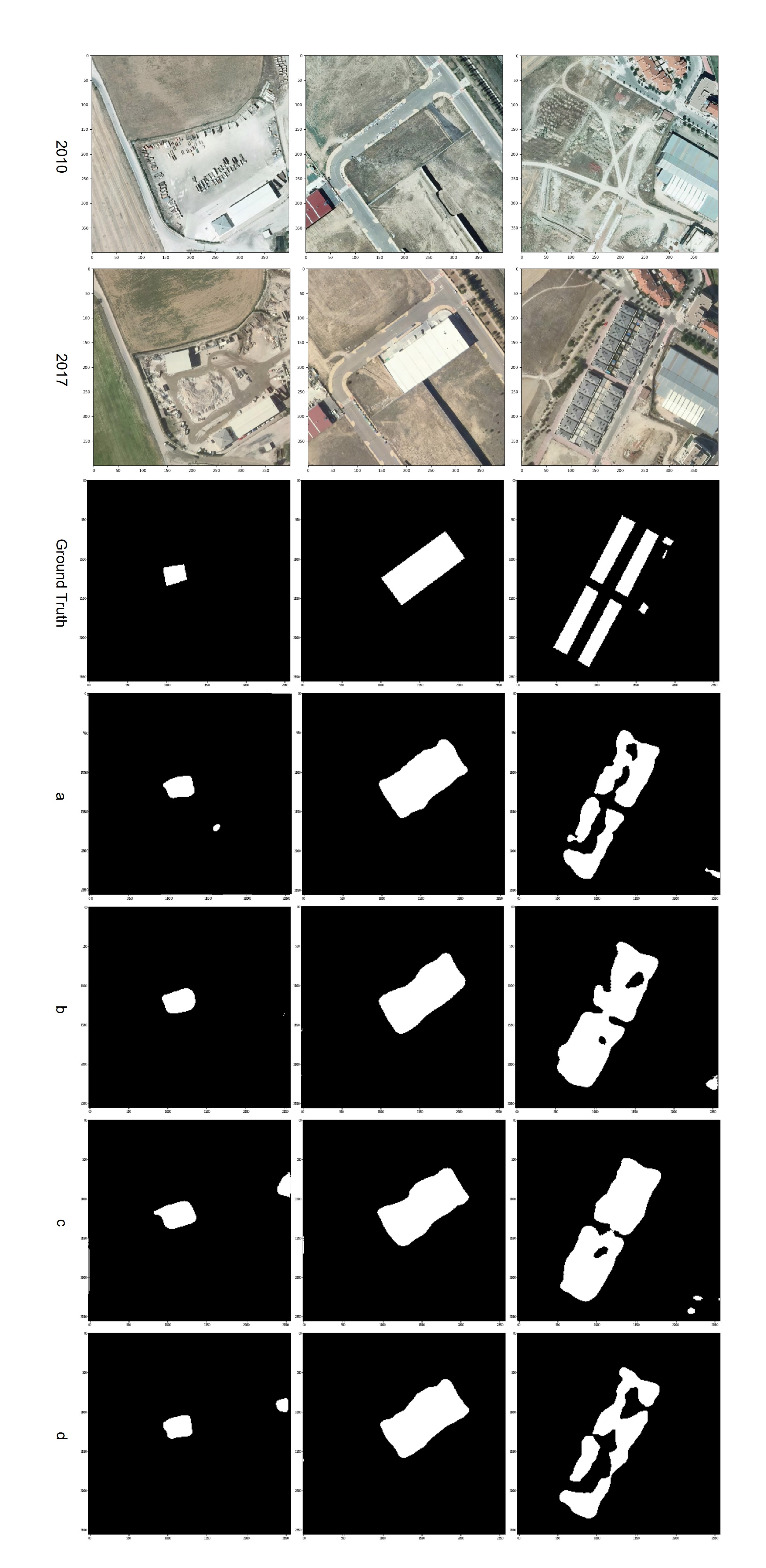}
    \caption{MTBIT 2D CD}
    \label{fig:mtbit2d_2}
    \end{subfigure}
    \hfill
    \begin{subfigure}[h]{0.48\textwidth}
      \centering
          \includegraphics[width=1\textwidth]{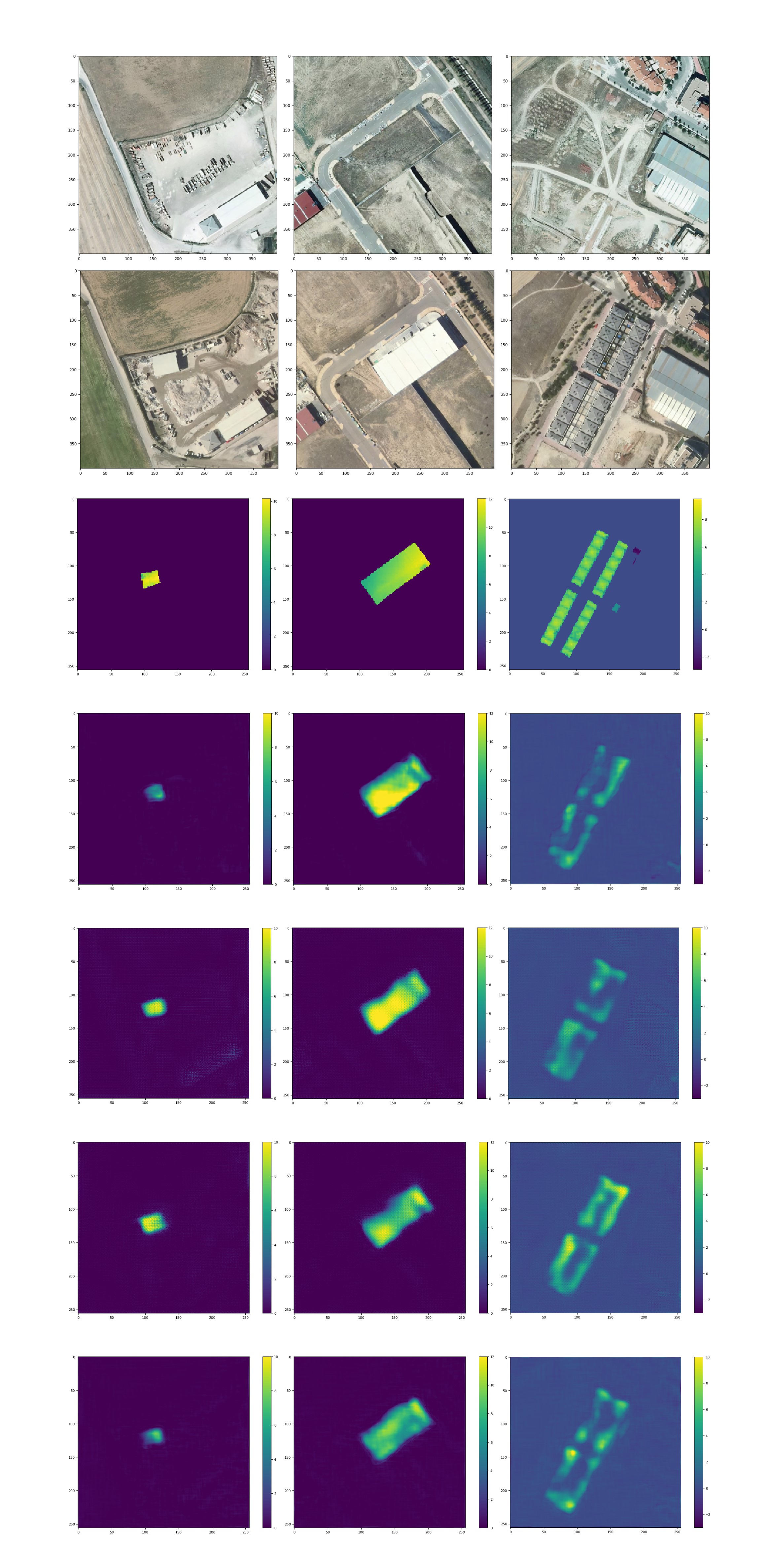}
    \caption{MTBIT 3D CD}
    \label{fig:mtbit3d_2}
      \end{subfigure}
\caption{Visualisation of 2D CD and 3D CD maps {obtained with the different} models described in Table \ref{tab:abl2}}
\label{fig:bit_2}
\end{figure}

In these ablation studies, we evaluated under which configuration MTBIT can achieve the best performance. {In particular, we tested} different features such as: type of upsampling (learnable or bilinear), concatenation or difference of features in the decoder output, number of attention heads in both the encoder and the decoder.
Specifically, we reported two sets of experiments: one concerning the sizes of the main components of the MTBIT architecture (see Table \ref{tab:abl1} and Figure \ref{fig:bit_1}) and one involving {specific} architecture features (see Table \ref{tab:abl2} and Figure \ref{fig:bit_2})

\begin{table}
\caption{Ablation studies assessing the behaviour of MTBIT with different sizes of the main components of the architecture.}
\label{tab:abl1}
\centering
\resizebox*{1\textwidth}{!}{
\begin{tabular}{|c|c|c|c|c|cc|cc|}
\hline
\multirow{2}{*}{\begin{tabular}[c]{@{}c@{}}Figure \ref{fig:bit_1}\end{tabular}} & \multirow{2}{*}{\begin{tabular}[c]{@{}c@{}}Encoder \\ depth\end{tabular}} & \multirow{2}{*}{\begin{tabular}[c]{@{}c@{}}Token \\ length\end{tabular}} & \multirow{2}{*}{\begin{tabular}[c]{@{}c@{}}Decoder\\ depth\end{tabular}} & \multirow{2}{*}{\begin{tabular}[c]{@{}c@{}}Decoder \\ dimension\\ head\end{tabular}} & \multicolumn{2}{c|}{\begin{tabular}[c]{@{}c@{}}2D CD\\ metrics\end{tabular}} & \multicolumn{2}{c|}{\begin{tabular}[c]{@{}c@{}}3D CD\\ metrics\end{tabular}} \\ \cline{6-9} 
 &  &  &  &  & \multicolumn{1}{c|}{\begin{tabular}[c]{@{}c@{}}F1\\ (\%)\end{tabular}} & \begin{tabular}[c]{@{}c@{}}IoU\\   (\%)\end{tabular} & \multicolumn{1}{c|}{\begin{tabular}[c]{@{}c@{}}RMSE\\   (m)\end{tabular}} & \begin{tabular}[c]{@{}c@{}}cRMSE\\   (m)\end{tabular} \\ \hline
a & 1 & 4 & 8 & 8 & \multicolumn{1}{c|}{63,02} & 46,00 & \multicolumn{1}{c|}{1,22} & 6,88 \\ \hline
b & 2 & 4 & 8 & 8 & \multicolumn{1}{c|}{62,05} & 44,98 & \multicolumn{1}{c|}{\textbf{1,17}} & 6,52 \\ \hline
c & 1 & 4 & 16 & 8 & \multicolumn{1}{c|}{\textbf{63,81}} & \textbf{46,86} & \multicolumn{1}{c|}{1,23} & 6,71 \\ \hline
d & 1 & 8 & 8 & 8 & \multicolumn{1}{c|}{60,98} & 43,86 & \multicolumn{1}{c|}{1,19} & 6,60 \\ \hline
e & 2 & 8 & 16 & 16 & \multicolumn{1}{c|}{59,82} & 42,68 & \multicolumn{1}{c|}{1,22} & 6,84 \\ \hline
f & 4 & 16 & 32 & 32 & \multicolumn{1}{c|}{61,61} & 44,52 & \multicolumn{1}{c|}{1,22} & 6,67 \\ \hline
g & 1 & 4 & 8 & 16 & \multicolumn{1}{c|}{62,15} & 45,09 & \multicolumn{1}{c|}{1,20} & \textbf{6,46} \\ \hline
\end{tabular}}
\end{table}

\begin{table}
\caption{Ablation studies assessing the behaviour of MTBIT with different features.}
\label{tab:abl2}
\centering
\resizebox*{1\textwidth}{!}{
\begin{tabular}{|c|c|c|c|cc|cc|}
\hline
\multirow{2}{*}{\begin{tabular}[c]{@{}c@{}}Figure \ref{fig:bit_2}\end{tabular}} & \multicolumn{1}{l|}{\multirow{2}{*}{\begin{tabular}[c]{@{}l@{}}Absolute value in \\ feature difference\end{tabular}}} & \multicolumn{1}{l|}{\multirow{2}{*}{Concat}} & \multicolumn{1}{l|}{\multirow{2}{*}{\begin{tabular}[c]{@{}l@{}}Learnable\\ upsampling\end{tabular}}} & \multicolumn{2}{c|}{\begin{tabular}[c]{@{}c@{}}2D CD\\ metrics\end{tabular}} & \multicolumn{2}{c|}{\begin{tabular}[c]{@{}c@{}}3D CD\\ metrics\end{tabular}} \\ \cline{5-8} 
 & \multicolumn{1}{l|}{} & \multicolumn{1}{l|}{} & \multicolumn{1}{l|}{} & \multicolumn{1}{c|}{\begin{tabular}[c]{@{}c@{}}F1\\ (\%)\end{tabular}} & \begin{tabular}[c]{@{}c@{}}IoU\\   (\%)\end{tabular} & \multicolumn{1}{c|}{\begin{tabular}[c]{@{}c@{}}RMSE\\   (m)\end{tabular}} & \begin{tabular}[c]{@{}c@{}}cRMSE\\   (m)\end{tabular} \\ \hline
a & Yes & No & No & \multicolumn{1}{c|}{\textbf{63,02}} & \textbf{46,00} & \multicolumn{1}{c|}{1,22} & 6,55 \\ \hline
b & No & Yes & No & \multicolumn{1}{c|}{60,91} & 43,79 & \multicolumn{1}{c|}{1,24} & 7,60 \\ \hline
c & No & No & Yes & \multicolumn{1}{c|}{61,11} & 44,00 & \multicolumn{1}{c|}{1,21} & 6,87 \\ \hline
d & No & No & No & \multicolumn{1}{c|}{62,15} & 45,09 & \multicolumn{1}{c|}{\textbf{1,20}} & \textbf{6,46} \\ \hline
\end{tabular}}
\end{table}

{In the first set of experiments, we started from a standard configuration \cite{bit} (see Table \ref{tab:abl1}-a) and we tried to increase the sizes of some of the components of the MTBIT architecture, to assess their effect on the performance of the network. 
As we can argue from the results, the architecture is fairly stable with respect to these changes, considering that the dataset is rather small. Particularly, we can observe that {larger} components of the network, especially when combined (see Table \ref{tab:abl1}-e and \ref{tab:abl1}-f), do not lead to better performances. This behaviour can be noted also in Figures \ref{fig:bit_1}-e and \ref{fig:bit_1}-f. {Moreover, Figure \ref{fig:att_map} shows the attention maps of two image examples. Once again, we can affirm that with a small dataset, $L = 4$ leads to a good performance of the network.}
\begin{figure}[h!]
    \centering
    \includegraphics[width=1\textwidth]{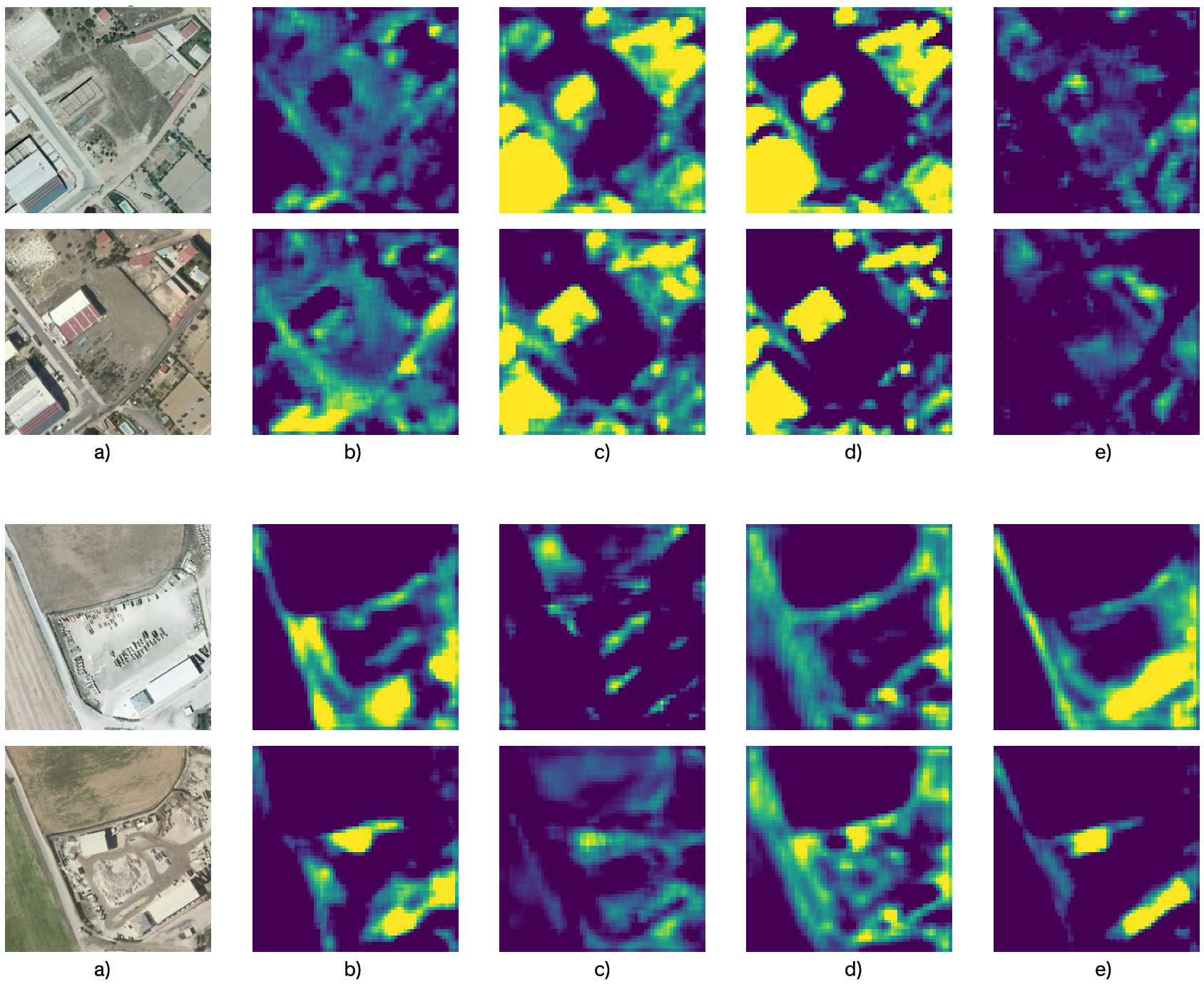}
    \caption{Visualisation of the attention maps for two bitemporal examples. Particularly, in a) the considered images are shown, from b) to e) the respective four attention maps are presented.}
    \label{fig:att_map}
\end{figure}
However, among all the tested configurations, we selected the one in which the decoder dimension head is 16 (Table \ref{tab:abl1}-g). In fact, under these conditions, MTBIT reached the best value of the cRMSE which is the mo{st} appropriate 3D CD metric, as previously described.}

{Concerning the modification of the network}, we found that a learnable upsampling, obtained through transposed convolutional modules, lead{s} to a worse performance with a higher number of parameters. 
Once again this {behaviour} is mainly due to an overfitting of the training data (Table \ref{tab:abl2}-c). Moreover, the intuition of concatenating the features, to {give} the model  {a} better understand{ing of the} elevation changes (with respect to LULC changes), was not confirmed by the experiments, even though the results are only slightly worse (Table \ref{tab:abl2}-b). {Nevertheless, we observed that to effectively understand the 3D changes and their sign (positive or negative), the network benefits by subtracting the features before the prediction heads, without considering their absolute values. Conversely, the absolute values work better for the 2D CD task, for which the sign is not representative of the phenomenon, as showed also in \cite{bit}.}

\subsubsection{Loss weights}
\label{sec:loss_w}

\begin{table}[h!]
\caption{Ablation studies assessing the behaviour of MTBIT with different weights.}
\label{tab:weights}
\centering
\begin{tabular}{|c|c|c|c|cc|cc|}
\hline
\multirow{2}{*}{Architecture} & 
\multirow{2}{*}{Figure \ref{fig:pesi}} & \multicolumn{1}{l|}{\multirow{2}{*}{$\alpha$}} & \multicolumn{1}{l|}{\multirow{2}{*}{$\beta$}} & \multicolumn{2}{c|}{\begin{tabular}[c]{@{}c@{}}2D CD\\ metrics\end{tabular}} & \multicolumn{2}{c|}{\begin{tabular}[c]{@{}c@{}}3D CD\\ metrics\end{tabular}} \\ \cline{5-8} 
 & & \multicolumn{1}{l|}{} & \multicolumn{1}{l|}{} & \multicolumn{1}{c|}{\begin{tabular}[c]{@{}c@{}}F1\\ (\%)\end{tabular}} & \begin{tabular}[c]{@{}c@{}}IoU\\   (\%)\end{tabular} & \multicolumn{1}{c|}{\begin{tabular}[c]{@{}c@{}}RMSE\\   (m)\end{tabular}} & \begin{tabular}[c]{@{}c@{}}cRMSE\\   (m)\end{tabular} \\ \hline
\multirow{7}{*}{MTBIT} &  a & 0 & 1 &  \multicolumn{1}{c|}{5,76} & 2,98 & \multicolumn{1}{c|}{1,24} & 6,93 \\ \cline{2-8} 
 &b & 1 & 0 &  \multicolumn{1}{c|}{19,40} & 10,74 & \multicolumn{1}{c|}{5,37} & 80,02 \\ \cline{2-8} 
 & c & 1 & 1 &  \multicolumn{1}{c|}{62,03} & 44,96 & \multicolumn{1}{c|}{1,22} & 6,75 \\ \cline{2-8} 
 & d & 1 & 3 & \multicolumn{1}{c|}{62,15} & 45,09 & \multicolumn{1}{c|}{1,20} & \textbf{6,46} \\ \cline{2-8} 
 &e & 3 & 1 & \multicolumn{1}{c|}{62,77} & 45,74 & \multicolumn{1}{c|}{1,21} & 6,80 \\ \cline{2-8} 
 & f& 1 & 5 &  \multicolumn{1}{c|}{62,71} & 45,68 & \multicolumn{1}{c|}{\textbf{1,19}} & 6,53 \\ \cline{2-8} 
 & g& 5 & 1 &  \multicolumn{1}{c|}{\textbf{63,65}} & \textbf{46,68} & \multicolumn{1}{c|}{1,24} & 6,83 \\ \hline
\end{tabular}
\end{table}

\begin{figure}[p]
    \centering
    \begin{subfigure}[h]{0.48\textwidth}
      \centering
          \includegraphics[width=1\textwidth]{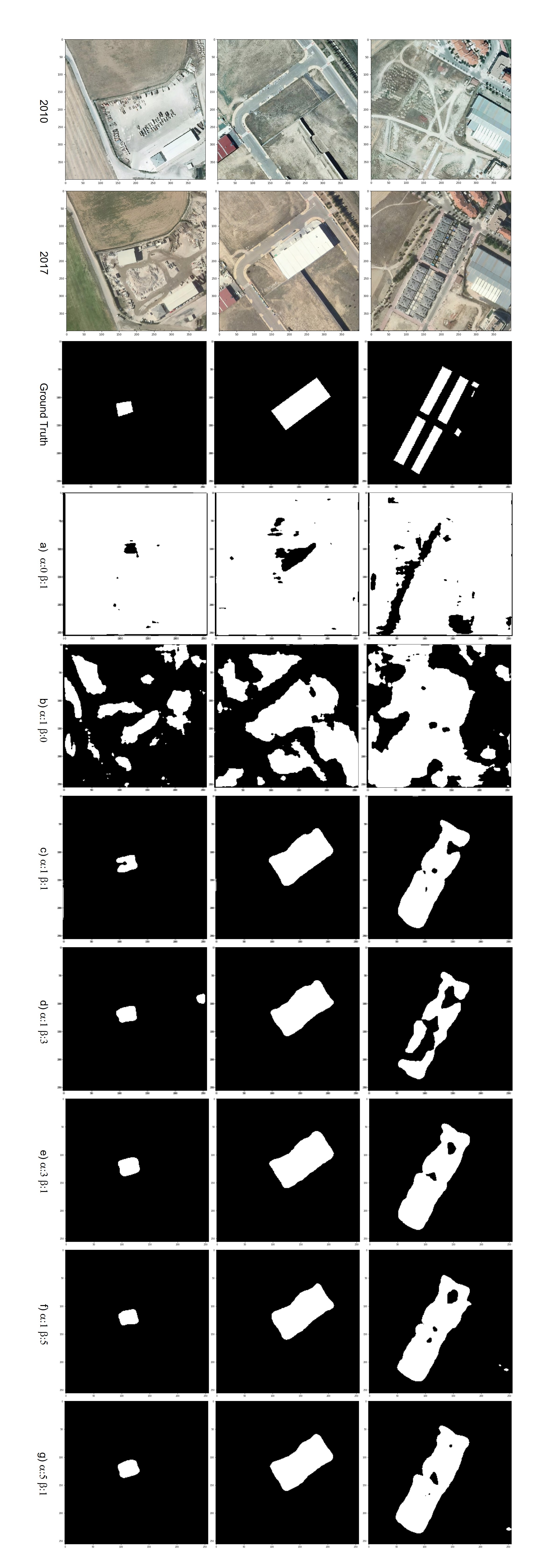}
    \caption{MTBIT 2D CD}
    \label{fig:bit_abl_2d}
    \end{subfigure}
    \hfill
    \begin{subfigure}[h]{0.48\textwidth}
      \centering
          \includegraphics[width=1\textwidth]{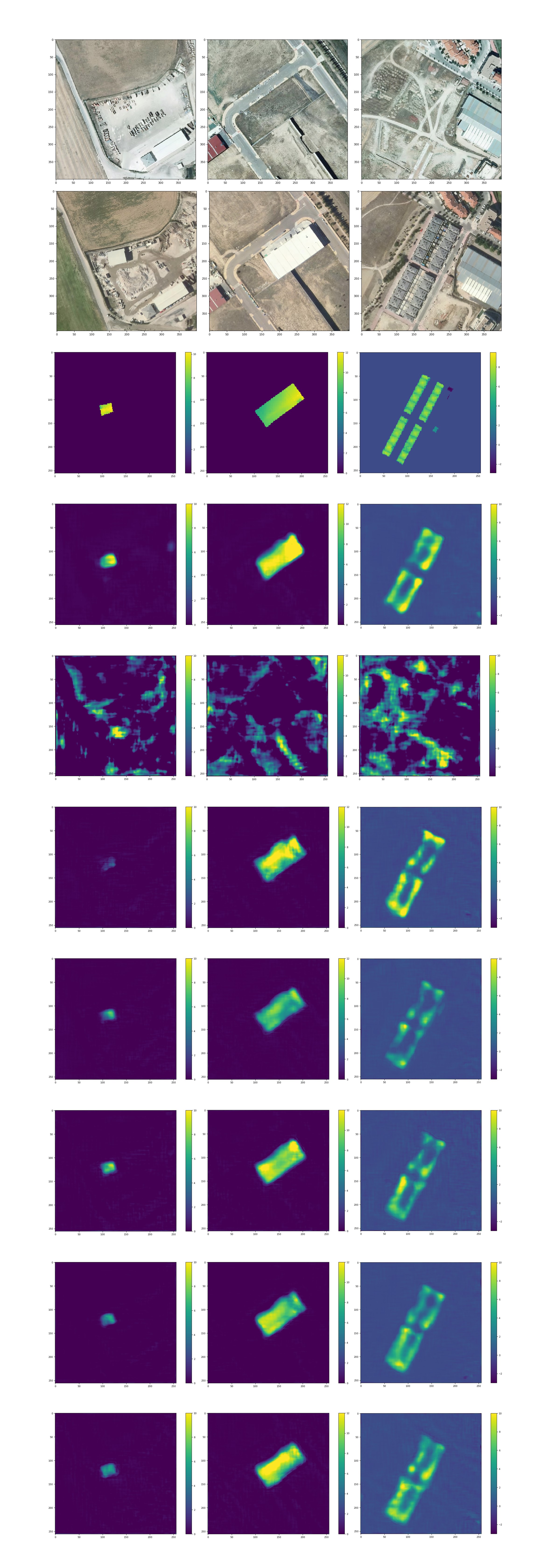}
    \caption{MTBIT 3D CD}
    \label{fig:bit_abl_3d}
      \end{subfigure}
\caption{Visualisation of 2D CD and 3D CD maps {obtained with the different} models described in Table \ref{tab:weights}}
\label{fig:pesi}
\end{figure}

Th{is} ablation study is aimed at analysing how MTBIT performs varying $\alpha$ and $\beta$, i.e. the weights of, respectively, the 2D loss and 3D loss {(Equation \eqref{eq:loss})}. In Table \ref{tab:weights} and Figure \ref{fig:pesi}, the quantitative and qualitative results are shown. 
Particularly, we can highlight {how} focusing exclusively on one task not only lead{s} to {a poor} performance on the other task, but do{es} not even {make it possible to} reach {the} best {performances on the considered task}. 
This is mainly due to the fact that the representations learned by the network to solve one task are {useful} also to solve the other task and viceversa. So, we {tested several combinations to find the optimal one}. As we can argue from the Table \ref{tab:weights} and Figures \ref{fig:pesi}, there is no{t} a{n evident} best combination {of the values of $\alpha$ and $\beta$} which {clearly} overperform{s} the others. 
The main intuition behind preferring higher values of $\beta$ is that {the} 2D CD task is easier to be solved and the 3D CD representations are more {effective} to solve {also} the 2D CD than viceversa. Concerning the magnitude of $\beta$, given $\alpha = 1$, the combination that reach{es the} most {balanced} results is with $\beta = 3$ (Figure \ref{fig:pesi}-d). Indeed, with lower (e.g., {the configuration with} $\beta = 1$ reaches {a} good performance on 2D CD, but does not achieve good results with respect to cRMSE) and higher (e.g., {the configuration with} $\beta = 5$ reaches unstable performance, where neither {the} 2D CD or {the} 3D CD metrics are the best in their categories) values of $\beta$, the MTBIT performances start to decrease.

\subsubsection{Encoders}
\label{sec:size}

\begin{table}
\caption{Ablation studies showing the behaviour of MTBIT under different encoders and strategies.}
\label{tab:size}
\resizebox*{1\textwidth}{!}{
\begin{tabular}{|c|c|c|c|c|cc|cc|}
\hline
\multirow{2}{*}{Figure: \ref{fig:encoder}}& \multirow{2}{*}{Encoder} & \multirow{2}{*}{\begin{tabular}[c]{@{}c@{}}Frozen\\ encoder\end{tabular}} & \multirow{2}{*}{\begin{tabular}[c]{@{}c@{}}Pretraining \\  strategy\end{tabular}} & \multicolumn{1}{c|}{\multirow{2}{*}{\begin{tabular}[c]{@{}l@{}}Trainable\\ parameters\\      (M)\end{tabular}}} & \multicolumn{2}{c|}{\begin{tabular}[c]{@{}c@{}}2D CD\\ metrics\end{tabular}} & \multicolumn{2}{c|}{\begin{tabular}[c]{@{}c@{}}3D CD\\ metrics\end{tabular}} \\ \cline{6-9} 
& &  &  & \multicolumn{1}{c|}{} & \multicolumn{1}{c|}{\begin{tabular}[c]{@{}c@{}}F1\\ (\%)\end{tabular}} & \begin{tabular}[c]{@{}c@{}}IoU\\   (\%)\end{tabular} & \multicolumn{1}{c|}{\begin{tabular}[c]{@{}c@{}}RMSE\\   (m)\end{tabular}} & \begin{tabular}[c]{@{}c@{}}cRMSE\\   (m)\end{tabular} \\
 \hline a &
ResNet18 & No & Random & 13,1 & \multicolumn{1}{c|}{49,24} & 32,66 & \multicolumn{1}{c|}{1,31} & 8,11 \\ \hline
b & ResNet18 & No & ImageNet & 13,1 & \multicolumn{1}{c|}{62,15} & 45,09 & \multicolumn{1}{c|}{1,20} & 6,46 \\ \hline
c & ResNet50 & No & Random & 42,9 & \multicolumn{1}{c|}{46,11} & 29,96 & \multicolumn{1}{c|}{1,40} & 8,16 \\ \hline
d & ResNet50 & No & ImageNet & 42,9 & \multicolumn{1}{c|}{\textbf{63,50}} & \textbf{46,52} & \multicolumn{1}{c|}{\textbf{1,15}} & \textbf{6,35} \\ \hline
e & ResNet18 & Yes & ImageNet & 1,4 & \multicolumn{1}{c|}{49,36} & 32,77 & \multicolumn{1}{c|}{1,28} & 7,11 \\ \hline
f & ResNet50 & Yes & ImageNet & 17,3 & \multicolumn{1}{c|}{56,72} & 39,58 & \multicolumn{1}{c|}{1,22} & 6,96 \\ \hline
\end{tabular}}
\end{table}

\begin{figure}[p]
    \centering
    \begin{subfigure}[h]{0.48\textwidth}
      \centering
          \includegraphics[width=1\textwidth]{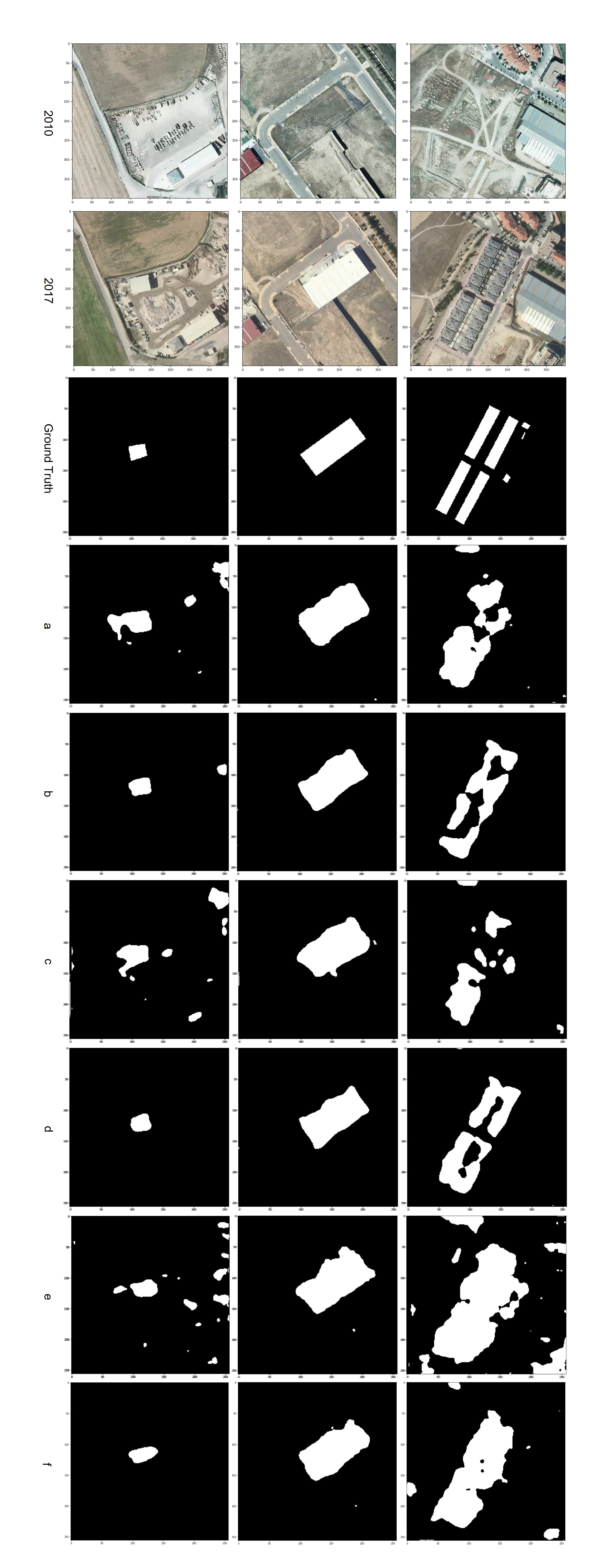}
    \caption{MTBIT 2D CD}
    \label{fig:bit2d_2}
    \end{subfigure}
    \hfill
    \begin{subfigure}[h]{0.48\textwidth}
      \centering
          \includegraphics[width=1\textwidth]{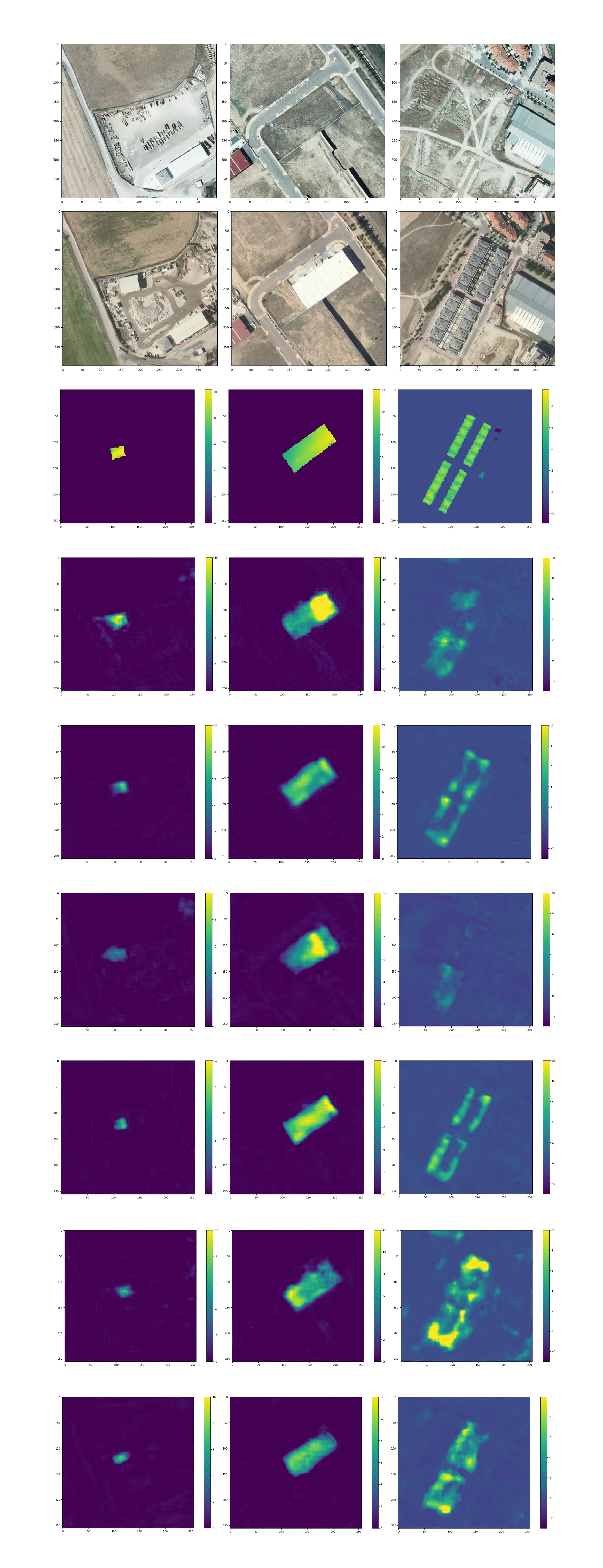}
    \caption{MTBIT 3D CD}
    \label{fig:bit3d_2}
      \end{subfigure}
\caption{Visualisation of 2D CD and 3D CD maps {obtained with the different} models described in Table \ref{tab:size}}
\label{fig:encoder}
\end{figure}

{In this set of ablation studies, we analyzed the behavior of MTBIT with different encoders and with different pretraining strategies. In particular, we tried to investigate how the performance of MTBIT changes with respect to the number of parameters and the initialization of the network. 
The results are shown in Figure \ref{fig:encoder} and in Table \ref{tab:size}.

First of all, as we expected, ResNet50 overcomes the ResNet18 performances.
However, with respect to the number of parameters, {which} is more than three times higher, MTBIT with ResNet50 encoder (42,9 M parameters) gains just 1.5\% in F1 and 11 cm in cRMSE with respect to MTBIT with ResNet18 encoder (13,1 M parameters), as Tables \ref{tab:size}-b and \ref{tab:size}-d show. 
For this reason, we selected the ResNet18 encoder model. 
Moreover, we can observe how initializating the encoder with pretrained Imagenet weights leads to a good performance of the network, both with a ResNet18 and a ResNet50 encoder, as it is reasonable to expect. In fact, even though the ImageNet features are rather different from the features of remote sensing images, the 3DCD dataset is too small to make the network learn rich features. In addition, we can observe that ResNet18 encoder reaches better values of the metrics than the ResNet50 encoder (e.g. 49.24\% vs 46.11\% for F1; 8.11 m vs 8.16 m for cRMSE) in the random weight initialization scenario. This is mainly due to the poor generalization capabilities of the larger network (MTBIT with ResNet50 encoder), that cause an overfitting of the training data. This is confirmed by the Figures \ref{fig:encoder}-a and \ref{fig:encoder}-c.
Regarding the experiments that involved freezing the encoder pretrained on ImageNet, we can observe that discrete performances are obtained. In general, we can affirm that freezing the encoder within such an architecture means to freeze the inputs of the semantic tokenizer. Since ImageNet features differ from the features of remote sensing images, we cannot expect good results. However, MTBIT with ResNet50 frozen encoder (17,3 M) loses 7\% in F1 and 0.5 m in cRMSE, emplyoing a number of parameters comparable to the one used by the benchmark MTBIT (13,1 M, see Table \ref{tab:comp}). Figures \ref{fig:encoder}-e and \ref{fig:encoder}-f show a distinct degradation, especially with respect to the 2D CD results. 
On the other hand, MTBIT with ResNet18 frozen encoder reaches similar performances to the ones obtained by the networks used for the comparison (Section \ref{sec:comparison}) with a similar number of parameters (see Table \ref{tab:comp}), such as ChangeFormer (1,8 M) and FC-EF (1,3 M)}.

\section{Conclusions}
\label{sec:conclusions}
\sloppy
In this paper, we presented a task that, to our knowledge, had not yet been investigated: {to infer elevation (3D) changes} from bitemporal optical images.
{The final aim was indeed to establish the basis for the development of DL algorithms able to automatically generate an elevation 3D CD map together with a standard 2D CD map, using only optical images captured at different epochs as input}.
{Specifically, we proposed MTBIT, a novel network belonging to the family of vision {Transformer}s, one of most used DL architectures, and a {modified} version of the 3DCD dataset, a freely available dataset designed precisely for this {twofold} task}.

In particular, the 3DCD dataset -- consisting of more than 450 {pairs of} images and DSMs together with the corresponding 2D and 3D CD maps -- {made it possible} to {address} the 3D CD problem in a direct and practical {way}. 

Furthermore, the MTBIT model, which extends to the new 3D CD task some of the methods already proposed for solving the 2D CD task, lays a solid foundation towards {the} solution of the 3D CD {problem}.  {In details, MTBIT achieves a metric accuracy (represented by the cRMSE) equal to 6.46 m, a value comparable to the one obtained by SUNet -- the most performant among the other tested architectures w.r.t. cRMSE -- but using only approximately one third of its parameters (13,1 M vs 35,7 M)}.

Nevertheless, much remains to be {investigated from} both the data and algorithmic {point of view}. 
First of all, an expansion of the dataset is already planned. This step is considered essential to enable a more effective application of those {architectures} that require a large amount of data to perform at their best. {Particularly, a fully-automated pipeline, able to annotate thousand of images, is currently under development. Moreover}, it is also important to work further on the {proposed network, which, despite the encouraging} results, {presents some} limitations, such as the difficult inference of extreme values. {In addition, other types of architectures, such as generative models (e.g. GANs or diffusion models), could be tested to solve {more} efficiently the 3D CD task in the future stages of the research.}

\bibliography{cas-refs}

\end{document}